\documentclass[%
 aip,
 amsmath,amssymb,
 reprint,%
]{revtex4-1}

\usepackage{graphicx}
\usepackage{dcolumn}
\usepackage{bm}

\usepackage[utf8]{inputenc}
\usepackage[T1]{fontenc}
\usepackage{mathptmx}
\newcommand\Alfven{Alfv\'en }
\newcommand\Alfvenic{Alfv\'enic }
\newcommand{\V}[1]{\mathbf{#1}} 
\newcommand{\T}[1]{{\tt #1}} 

\draft 

\begin{document}

\title[Electron Landau Damping of KAWs]{Electron Landau Damping of Kinetic Alfv\'en Waves in Simulated Magnetosheath Turbulence}

\author{Sarah A. Horvath}
\email[E-mail: ]{sarah-horvath@uiowa.edu}

\author{Gregory G. Howes}
\email[E-mail: ]{gregory-howes@uiowa.edu}

\author{Andrew J. McCubbin}
\email[E-mail: ]{andrew-mccubbin@uiowa.edu}
\affiliation{Department of Physics and Astronomy, University of Iowa, Iowa City, Iowa 52242, USA}

\date{\today}

\begin{abstract}

Turbulence is thought to play a role in the heating of the solar wind plasma, though many questions remain to be solved regarding the exact nature of the mechanisms driving this process in the heliosphere. In particular, the physics of the collisionless interactions between particles and turbulent electromagnetic fields in the kinetic dissipation range of the turbulent cascade remains incompletely understood. A recent analysis of an interval of \emph{Magnetosphere Multiscale} (\emph{MMS}) observations has used the field-particle correlation technique to demonstrate that electron Landau damping is involved in the dissipation of turbulence in the Earth's magnetosheath. Motivated by this discovery, we perform a high-resolution gyrokinetic numerical simulation of the turbulence in the \emph{MMS} interval to investigate the role of electron Landau damping in the dissipation of turbulent energy. We employ the field-particle correlation technique on our simulation data, compare our results to the known velocity-space signatures of Landau damping outside the dissipation range, and evaluate the net electron energization. We find qualitative agreement between the numerical and observational results for some key aspects of the energization and speculate on the nature of disagreements in light of experimental factors, such as differences in resolution, and of developing insights into the nature of field-particle interactions in the presence of dispersive kinetic Alfv\'en waves.

\end{abstract}


\maketitle 

\section{\label{intro} Introduction }

The mechanisms that govern the dissipation of turbulent energy in space and astrophysical plasmas, where the energy of turbulent fields and flows is transferred to the plasma particles as heat or some other non-thermal form of energization, remain poorly understood, representing a key challenge at the frontier of heliophysics. Many physical processes have been put forward as potential contributors to solar wind particle energization, including Landau damping, transit-time damping, cyclotron damping \citep{Barnes:1966,TenBarge:2013a,Schekochihin:2009}, stochastic ion heating, magnetic pumping \citep{Chen:2001,Lichko:2017}, and dissipation in coherent structures \citep{Dmitruk:2004,Karimabadi:2013}. In this work, we focus in particular on the contribution of Landau damping by electrons. Landau damping is a resonant wave-particle interaction in which energy is transferred collisionlessly from the electromagnetic fields to non-thermal energy contained in fluctuations in velocity-space. Subsequent processes transfer this free energy in the particle velocity distribution to sufficiently small velocity-space scales that arbitrarily weak collisions are sufficient to increase the entropy, and thus lead to irreversible thermodynamic heating of the plasma \citep{Howes:2006,Schekochihin:2009,Tatsuno:2009,Howes:2017}. 

Landau damping by ions in a turbulent plasma has been studied numerically in the past and is fairly well understood \citep{Klein:2017,Howes:2017b,Howes:2018,Klein:2020}. Electron dynamics, however, are both less well understood and more challenging to study due to the smaller spatial scales and higher frequencies involved. When solar wind turbulence cascades beyond the ion scales at the ion Larmor radius, the cascade is thought to transition from the magnetohydrodynamic (MHD) inertial range into the kinetic dissipation range \citep{Howes:2008b,Howes:2015,Kiyani:2015}. Here, kinetic theory is required since the MHD approximation is no longer valid, and the MHD Alfv\'en waves cascade into dispersive kinetic Alfv\'en waves (KAWs) \citep{Howes:2011b}. The dispersive nature of the KAWs in this region introduces new and relatively unstudied complexities into processes such as Landau damping.

Modern instrumentation and computer capabilities make research on kinetic-scale solar wind phenomenon possible, though it remains challenging. A study by Chen et al. (2019)\citep{Chen:2019} (hereafter CKH19) is a successful example of using observational data to study kinetic-scale processes in the heliosphere, and led to the first direct evidence for electron Landau damping in the solar wind \citep{Chen:2019}. In that work, the field-particle correlation technique \citep{Klein:2016,Howes:2017,Klein:2017,Howes:2018} was applied to electrons in the Earth's magnetosheath using a data interval from a \emph{Magnetospheric Multiscale} (\emph{MMS}) probe.

Another challenge that must be overcome in order to utilize \emph{in situ} observations, in addition to the need for high-resolution in both time and velocity-space, is the inherent single-point nature of spacecraft data. For many previous studies of wave-particle interactions using single-point data, including the CKH19 study and studies of simulated plasmas, this challenge has successfully been addressed by using the field-particle correlation technique. This technique using single-point velocity distribution and electromagnetic field data to analyze energy transfer in solar wind turbulence \citep{Klein:2016}. 

Motivated by the identification of electron Landau damping in the magnetosheath, we choose to further explore the nature of this damping mechanism by modeling the turbulent magnetosheath plasma interval of CKH19 using the gyrokinetic code \T{AstroGK} and applying the field-particle correlation technique to the synthetic data we produce. This work is a first step toward understanding the effects of the dispersive nature of KAWs on electron energization by a resonant mechanism. By furthering knowledge of the smallest-scale regions of the solar wind turbulent cascade, we hope to inform big picture questions such as the identification of the physical mechanisms responsible for the turbulent heating of particles in the heliosphere. 

In Sec. \ref{sec:FPC}, we briefly summarize the field-particle correlation technique, which has been described in detail in previous works \citep{Howes:2017, Klein:2017, Klein:2016}. An overview of particle energization by the dissipation of turbulence is given in Sec. \ref{sec:PEDT}. The \T{AstroGK} simulation setup and the synthetic magnetosheath plasma it produced are described in Sec. \ref{sec:AGKS}, and Sec. \ref{sec:FPCA} describes the results of field-particle correlation technique analysis on this data. In Sec. \ref{sec:D} we discuss the similarities and differences between our result and the observational result of CKH19, speculate on the cause of discrepancies, and consider how current and future work may contribute to a deeper understanding of this topic.

\section{\label{sec:FPC} The Field-Particle Correlation Technique }

The Maxwell-Boltzmann equations govern the dynamics of the magnetized, weakly collisional, heliospheric plasma. Maxwell's equations describe the evolution of the fields, and the Boltzmann equation describes the evolution of the six-dimensional distribution function for a plasma species, $f_s({\bf r}, {\bf v}, t)$. On the timescale of the collisionless transfer of energy between the electromagnetic fields and the plasma particles, which governs the crucial first step in the dissipation of turbulent energy \citep{Howes:2008c,Schekochihin:2009}, the collisional term in the Boltzmann equation is unimportant \citep{Howes:2015,Howes:2015b}.  Therefore, for the purpose of understanding the electron energization, we set the collision operator in the Boltzmann equation to zero, yielding the Vlasov equation (in cgs units): 
\begin{equation} \frac{\partial f_s}{\partial t} +{\bf v} \cdot {\bf \nabla} f_s + \frac{q_s}{m_s}\left[ {\bf E} + \frac{{\bf v} \times {\bf B}}{c}  \right] \cdot  \frac{\partial f_s}{\partial {\bf v}} = 0.
\end{equation}
We define a quantity called the phase-space energy density,
\begin{equation} w_s({\bf r}, {\bf v},t) = \frac{m_s v^2}{2} f_s({\bf r}, {\bf v}, t),
\end{equation} noting that the Vlasov equation multiplied by $m_s v^2/ 2$ gives the rate of change of this quantity: 
\begin{equation} \frac{\partial w_s}{\partial t} =  - {\bf v} \cdot {\bf \nabla} w_s - q_s \frac{v^2}{2} {\bf E} \cdot \frac{\partial f_s}{\partial {\bf v}} - \frac{q_s}{c}\frac{v^2}{2} ({\bf v} \times {\bf B}) \cdot  \frac{\partial f_s}{\partial {\bf v}}.
\end{equation}
If we integrate equation (3) over velocity-space, we see that the third term on the right-hand side goes to zero after integrating by parts. Unsurprisingly, the magnetic field cannot contribute to a net energy change as it does no work on the particles. If we integrate over all volume, the first term on the right-hand side also has a null contribution (for periodic or infinite boundary conditions). Thus, the second term must be responsible for any net energy changes the particles experience \citep{Klein:2016, Howes:2017}. This contributing term depends only on the electric field and the velocity derivative of the distribution function, both of which can be obtained from single-point measurements at some point ${\bf r_0}$. 

By correlating the single-point measurements of ${\bf E}$ and $\partial f_s/\partial {\bf v}$ over a sufficiently long time interval $\tau$, we obtain the net rate of particle energization at ${\bf r_0}$. The parallel correlation (note that it is the parallel component of the electric field, $E_{\parallel}$, that contributes to Landau damping) is: 
\begin{equation}  C_{E_{\parallel}}({\bf v}, t, \tau) = C\left( -q_s \frac{v_{\parallel}^2}{2}\frac{\partial f_s({\bf r_0}, {\bf v}, t)}{\partial v_{\parallel}}, E_{\parallel}({\bf r_0}, t) \right),
\end{equation} 
where $C$ is the unnormalized correlation defined by: 
\begin{equation} C(A,B) = \frac{1}{N} \sum_{j=i}^{i+N-1}A_j B_j ~. 
\end{equation} 
The above summation is essentially a time average over $N$ consecutive time steps, where we define the \emph{correlation interval} as $ \tau = N\Delta t$. The correlation interval is chosen to span several Alfv\'en wave periods, such that the time average causes large-scale oscillatory energy transfer to predominantly cancel out. Any small-scale secular energy transfer from fields to particles that was previously obscured---often this secular transfer of energy is an order of magnitude or more smaller than the oscillatory transfer \citep{Howes:2017}---is then observable. In addition to uncovering the magnitude of the net energy transfer, this technique preserves the velocity-space location of the energization, revealing a unique \emph{velocity-space signature} for the physical mechanism at work \citep{Klein:2016,Howes:2017,Klein:2020}. Landau damping, the process of interest here, has been shown to produce a bipolar signature in gyrotropic velocity-space ($v_{\parallel}, v_{\perp}$) that is consistent with theory regarding wave-particle interactions due to Landau damping. 

\section{\label{sec:PEDT} Particle Energization by Dissipation of Turbulence}

\subsection{Overview of Turbulent Particle Energization}
The turbulent cascade in the solar wind is comprised of three main regions: (i) the \emph{energy containing range}, (ii) the \emph{inertial range}, and (iii) the \emph{dissipation range} \citep{Tu:1995,Bruno:2005,Kiyani:2015,Howes:2015b}. The energy containing range holds the largest turbulent structures on length scales $L \gtrsim 10^{6}$ km, scales so large that the fluctuations have not had time to interact nonlinearly (and thereby contribute to the turbulent cascade of energy) within the time it has taken for the solar wind to travel from the Sun to the point of observation (at $\sim 1$ AU). At slightly smaller scales the inertial range begins, in which the \Alfvenic turbulent dynamics are well modeled by reduced MHD \citep{Schekochihin:2009}. The energy cascade in this region, dominated by Alfv\'en waves, is self-similar and scale-invariant in the manner of Komolgorov hydrodynamic turbulence \citep{Kolmogorov:1991}, and follows an anisotropic power law spectrum either $\propto k_\perp^{-5/3}$ (consistent with Goldreich and Schridar's 1995 (GS95) analysis of anisotropic MHD turbulence) \citep{Goldreich:1995, Kiyani:2015} or $\propto k_\perp^{-3/2}$ (consistent with Boldyrev's 2006 theory for dynamic alignment within anisotropic MHD turbulence) \citep{Boldyrev:2006}. This anisotropic, \Alfvenic cascade continues without dissipation through the inertial range until it reaches perpendicular length scales that are on the order of the ion gyroradius \citep{Schekochihin:2009}, $\rho_i$, where the dissipation range begins. At these scales, the MHD fluid approximation is no longer valid and the plasma must be modeled kinetically \citep{Howes:2008c,Schekochihin:2009,Kiyani:2015}. In the dissipation range, the individual ions, and later electrons, interact collisionlessly with the turbulent structures and the cascade spectrum steepens due to the dispersive nature of the plasma response at sub-ion scales and as the particles remove energy from the turbulence and generate non-thermal energy in their velocity distributions. Ultimately, this non-thermal energy can be transferred via a phase-space cascade to small scales in both configuration and velocity-space, where arbitrarily weak collisions can thermalize that energy, increasing the entropy and thereby irreversibly heating the plasma species \citep{Howes:2008c, Schekochihin:2009,Tatsuno:2009}.

In this work, we specifically focus on the energization of electrons within the dissipation range. Due to the  large ion-to-electron temperature ratio in the \emph{MMS} interval,  $T_i/T_e = 9$, the scales corresponding to the electron gyroradius, $k_{\perp} \rho_e \sim 1 $, are in terms of the ion gyroradius approximately $k_{\perp} \rho_i \sim 128$. Resolving this large separation of scales is extremely challenging for a kinetic simulation, so we choose to drive our cascade well into the dissipation range at $k_{\perp 0} \rho_i = 8$. Based on expectations from linear kinetic theory, the collisionless damping rate remains small at ion scales ( $\gamma/\omega \leq 10^{-2}$, see Appendix A). Thus, we expect little of the turbulent energy to be transferred to the ions, and essentially all of the turbulent energy to cascade to sub-ion scales, be transferred to the electrons by collisionless mechanisms, and ultimately lead to electron heating. Therefore, we require a fully kinetic simulation with high resolution in the dissipation range in order to accurately model the \emph{MMS} interval.

\subsection{Field-Particle Correlation Analysis of Electron Landau Damping}
\label{sec:diagram}

Electron Landau damping in a turbulent plasma has not previously been studied in detail by the field-particle correlation technique. Though some important differences are expected to exist between electron and ion Landau damping, the better-studied example of ions gives important insight into what we can expect for electrons. In ion Landau damping, one finds a bipolar velocity-space signature of energization using the field-particle correlation technique \citep{Klein:2017,Klein:2016,Howes:2017,Howes:2018}. 

\begin{figure}[t]
\includegraphics[bb=35 5 370 306, width=0.48\textwidth]{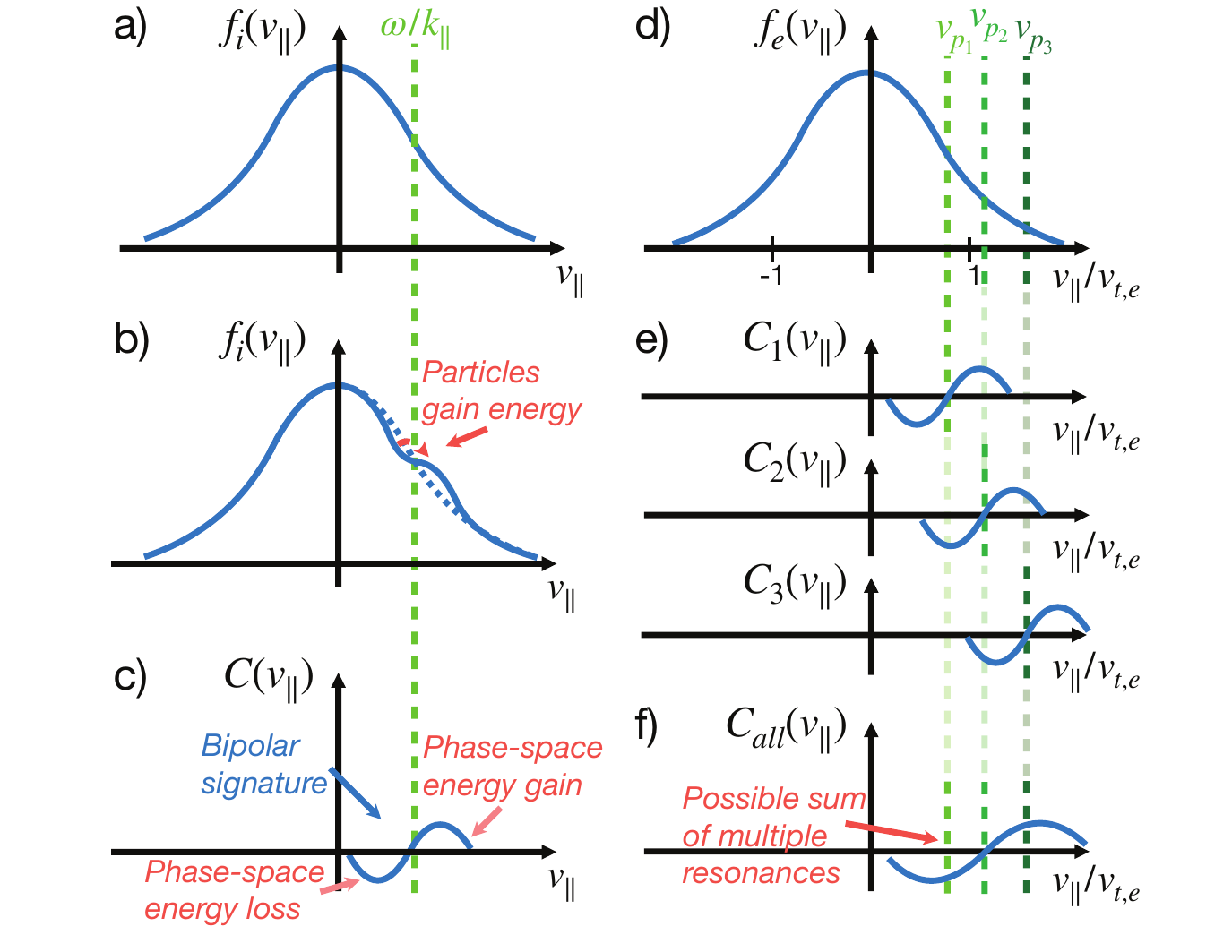}
\caption{\label{fig:LDdiagramFull} \textit{(Left)} Illustration of ion Landau damping: (a) a Maxwellian ion velocity distribution function and Alfv\'en wave phase velocity, $\omega/k_\parallel$; (b) velocity distribution changes due to Ion Landau damping of the Alfv\'en wave; (c) resulting phase-space signature of the correlation, $C(v_{\parallel})$, after field-particle correlation analysis.  \textit{(Right)} Speculative illustration of electron Landau damping by dispersive KAWs: (d) a Maxwellian electron velocity distribution and three KAWs of varying phase velocity; (e) the individual field-particle correlation signatures; (f) a possible resultant signature superposition.}  
\end{figure}

The left hand column of Fig.~\ref{fig:LDdiagramFull} illustrates the physics behind this bipolar signature. In panel (a), we present a Maxwellian velocity distribution function of ions moving along the mean magnetic field, $f_i(v_{\parallel})$. The resonant denominator in kinetic theory that governs the Landau resonance has the form $\omega - k_\parallel v_\parallel = 0$, so ions with parallel velocities near the phase velocity of the \Alfven wave, $v_\parallel \simeq \omega/k_{\parallel}$, will interact resonantly with the Alfv\'en wave at this phase velocity. Note that we adopt the convention that $\omega >0$, so that the sign of $k_\parallel$ determines the direction of propagation of the \Alfven wave: $\omega/k_{\parallel}>0$ corresponds to an \Alfven wave propagating up the magnetic field, and $\omega/k_{\parallel}<0$ to one propagating down the field.  Therefore, the \Alfven wave in Fig.~\ref{fig:LDdiagramFull}(a) is propagating up the field, and interacts resonantly with ions moving in the same direction with $v_\parallel >0$.

The net effect of this resonant interaction, as shown in Fig.~\ref{fig:LDdiagramFull}(b), is that ions initially moving at velocities slightly greater than the phase velocity, $v_\parallel > \omega/k_{\parallel}$, will lose energy to the wave; those initially moving at velocities slightly lower than the phase velocity, $v_\parallel < \omega/k_{\parallel}$, will gain energy from it. If the slope of the distribution function is negative at the resonance, the number of particles initially moving slower than the resonant velocity is greater than the number initially moving faster, resulting in a net loss of particles with $v_\parallel < \omega/k_{\parallel}$ and a net gain of particles with $v_\parallel > \omega/k_{\parallel}$. Effectively, Landau damping results in a quasilinear flattening of the ion distribution function at the resonant velocity, as shown in panel (b). This perturbation to the velocity distribution leads to an overall increase in phase-space energy density, $w(v_{\parallel}) = m v_{\parallel}^2 f_i(v_{\parallel})/2$, because of its weighting by $v_\parallel^2$. The corresponding rate of change of phase-space energy density as a function of $v_\parallel$, determined by computing the reduced parallel field-particle correlation $C_{E_\parallel} (v_\parallel)$, yields the characteristic bipolar velocity-space signature of ion Landau damping, shown in panel (c): a loss of phase-space energy density just below the resonance at $v_\parallel \simeq \omega/k_{\parallel}$, and a gain of phase-space energy density just above this resonance. The velocity-space signature is a visual representation of the gain and loss of energy in phase-space, and its quantitative and qualitative characteristics enable one to distinguish different processes of particle energization \citep{Klein:2017,Howes:2017b,Howes:2018,Klein:2020}.

For ions, both MHD and kinetic Alfv\'en waves are relatively nondispersive over the narrow range of $k_{\perp} \rho_i$ at which the ion damping rate is significant. Therefore, a single bipolar velocity-space signature of Landau damping is typically the only feature observed in the ion correlation \citep{Klein:2017,Howes:2017b,Howes:2018,Klein:2020}, as illustrated in panel (c). For electrons, however, KAWs are dispersive, resulting in a range of phase velocities over which the electrons may be energized by waves with different values of $k_{\perp} \rho_i$ (see Appendix A). Therefore, the dispersive nature of these KAWs raises an important question: \emph{What velocity-space signature will the field-particle correlation technique produce for electron energization by Landau damping in a plasma exhibiting a broadband turbulent cascade?}

Fundamentally, the physical process of electron Landau damping is exactly analogous to ion Landau damping; however, the dispersive nature of the KAWs may result in multiple wave modes interacting simultaneously with the plasma particles at various resonant velocities. In Fig.~\ref{fig:LDdiagramFull}(d)-(f), we illustrate a possible scenario and outcome of the phase-space signature for dispersive KAWs. At the top, we see a Maxwellian electron velocity distribution and the phase velocities $v_{p_1}$, $v_{p_2}$, and $v_{p_3}$ of three linearly superposed KAWs with different perpendicular scales $k_\perp \rho_i > 1$.
At each of these resonant velocities electron Landau damping may occur, exactly analogous to panel (b), resulting in three phase-space correlation signatures with each centered around its respective resonance, shown in panel (e). All three waves in this example alter the phase-space energy density simultaneously at the spatial point of observation. One possibility for the resulting signature, $C_{all}(v_{\parallel})$, is a simple superposition of all individual signatures, yielding a broadened bipolar signature as shown in panel (f). For the typically monotonically decreasing magnetic energy spectrum in the dissipation range of solar wind turbulence, however, one may need to weigh each of these predicted overlapping bipolar signatures by the amplitude of the fluctuations at perpendicular wavenumber $k_\perp \rho_i$ as well as the damping rate of KAWs at that wavenumber. Futhermore, KAWs of different $k_\perp \rho_i$ may pass through the point of observation at different times, leading to a much more complicated structure in  $C_{E_\parallel} (v_\parallel)$ than the simple broadened signature shown in Fig.~\ref{fig:LDdiagramFull}(f).

A key goal of this work is to determine numerically the qualitative appearance of the velocity-space signature of electron Landau damping in a turbulent plasma. Specifically, we choose to tackle this goal using a gyrokinetic simulation of kinetic-scale turbulence corresponding to the interval of \emph{MMS} magnetosheath turbulence studied in CKH19.

\section{\label{sec:AGKS} Simulation Setup}
\subsection{AstroGK Parameters}
The turbulence and plasma parameters corresponding to the 70~s \emph{MMS} interval used in CKH19 were simulated using the Astrophysical Gyrokinetics Code (\T{AstroGK}) \citep{Numata:2010}, which has had a successful history of modeling astrophysical plasmas since its development a decade ago \citep{Howes:2008,Howes:2011,TenBarge:2013b,Howes:2011}. Gyrokinetics is a limit of kinetic plasma theory that is relevant when three conditions are fulfilled: (i) weak coupling, $n_{0e}\lambda^3_{De} \gg 1$, (ii) strong magnetization, $\rho_i = \frac{v_{th_i}}{\Omega_i} \ll L$, where $L$ is the length scale of gradients in the plasma equilibrium, and (iii) low frequencies, $\omega \ll \Omega_i $ \citep{Howes:2006,Schekochihin:2009}. For the low-frequency, \Alfvenic turbulence observed in many heliospheric environments, these limits are typically satisfied \citep{Howes:2008b}.  The low-frequency limit enables all quantities to be averaged over the gyromotion of the particles, which effectively reduces the phase-space from six to five dimensions: three in space and two in velocity. The velocity dimensions are usefully described in relation to the mean magnetic field (note that we align our coordinates such that ${\bf B_0} = B_0 {\bf \hat{z}}$): $v_{\parallel}$ describes particle motion along ${\bf B_0}$, and $v_{\perp}$ describes the perpendicular motion.

\begin{figure}[t]
\includegraphics[width=0.19\textwidth, height=0.25\textwidth]{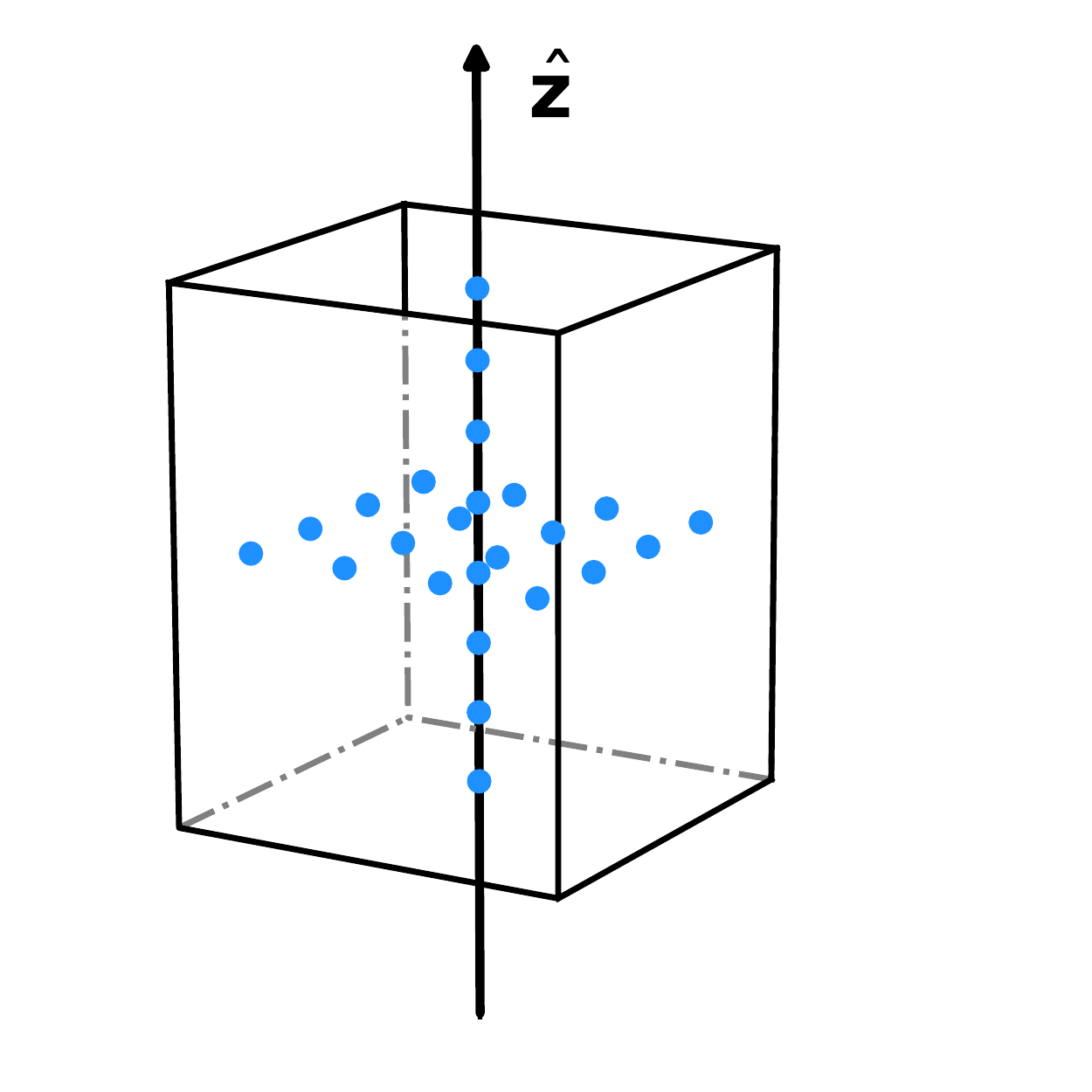}
\caption{\label{fig:simbox} Illustration of the simulation box, indicating the orientation with respect to ${\bf \hat{z}}$ and the intended distribution of the $24$ probes.} 
\end{figure}

In physical space, the simulated plasma occupies a rectangular box, elongated in the direction of the mean magnetic field, $L_\parallel \gg L_\perp$, to accommodate the anisotropic nature of the strong turbulent cascade, with fluctuations that satisfy $k_\perp \gg k_\parallel$ \citep{Goldreich:1995,Cho:2002,Sahraoui:2010} . Twenty-four probe points are distributed throughout the box, sixteen in the $xy$-plane at $z=0$, and the remaining eight along the $z$-axis, as illustrated in Fig.\ref{fig:simbox}. Note that due to an error in the diagnostic outputs of this particular simulation, the xy-positions of all twenty-four probe points have been shifted from their intended locations by a factor of $1/8$. It is important to note that the integrity of the simulation itself was not affected, and that single-point field-particle correlations are unhindered. The simulation volume that we are able to probe is simply reduced in two of its intended dimensions, but our  results still provide meaningful insight into electron Landau damping in the region probed.

The plasma parameters of this simulation are chosen to match the \emph{MMS} observation: ion plasma beta $\beta_i = v_{ti}^2/v_A^2 = 0.8$, ion-to-electron temperature ratio $T_i/T_e = 9$, and a realistic proton-to-electron mass ratio $m_i/m_e = 1836$. The dimensions of the simulation are $(n_x, n_y, n_z, n_\lambda, n_\varepsilon, n_s) = (64, 64, 32, 128, 32, 2)$, where we note that pitch angle and energy are given by $\lambda = v^2_{\perp}/v^2$ and $\varepsilon = v^2_{\perp} + v^2_{\parallel}$, respectively. The spatial domain of the simulation is an elongated box of dimension $L_{\perp}^2 \times L_{\parallel}$, with periodic boundary conditions. We set the relationship between the simulation box size and the driving scale wavenumbers to be $k_{\perp 0} = 2 \pi/ L_{\perp}$, $k_{\parallel 0} = 2 \pi/ L_{\parallel}$. In dimensionless units, where the perpendicular wavenumber is normalized by particle Larmor radius, $k_{\perp}\rho_i$ is fully resolved over the range $ 8 \le k_{\perp}\rho_i \le 168 $, or $k_{\perp}\rho_e$ over the range $ 0.062 \le k_{\perp}\rho_e \le 1.31 $. This range of resolved wave modes satisfies our goal of simulating the kinetic Alfv\'en wave cascade and its dissipation onto the electrons. The complex frequency of KAWs at the domain scale $k_{\perp 0}\rho_i = 8$ is $(\overline{\omega}_0, \overline{\gamma}_0) = (4.996, -0.257)$, where the bar indicates normalization by the Alfv\'en velocity, $\overline{\omega}_0 = \omega/k_{\parallel 0} v_A$ . We define the period of domain scale waves as $T_0 = 2 \pi/\omega_0$.  Ion and electron collisionalities are set to $\nu_i=\nu_e=8.94\times 10^{-3} k_\parallel v_A$, leading to weakly collisional conditions.

\subsection{Global Evolution}

We inject energy into the turbulent cascade by driving four perpendicularly polarized, counterpropagating KAWs at $k_{\perp 0} \rho_i=8$ using an oscillating Langevin antenna \citep{TenBarge:2014} with real frequency $\overline{\omega}_{ant}=4.5$ and decorrelation rate $\overline{\gamma}_{ant} =  -3.0$. The amplitude of the driven waves is set to obtain a strongly turbulent, critically balanced cascade. This technique models energy injection at the domain scale due to nonlinear interactions among turbulent fluctuations slightly larger than the simulation domain that transfer their energy nonlinearly to smaller scales. This driving method results in a steady state, strong turbulent cascade by simulation time $t/T_0= 0.3494$, consistent with earlier simulations of kinetic Alfv\'en wave turbulence \citep{Howes:2017c}. The turbulence remains in steady state until we end the simulation at $t/T_0 = 8.5834$. To ensure that we are analyzing only fully-developed turbulence, we restrict our interval to $t/T_0 = [0.4208, 8.5834]$. Note that this time interval spans $8.16$ outer-scale kinetic Alfv\'en waves periods (corresponding to $k_{\perp 0}\rho_i = 8$); $27.74$ wave periods at the smallest fully resolved scales ($k_{\perp}\rho_i = 168$); and $32.67$ of the highest frequency KAWs ($k_{\perp}\rho_i = 70.85$). The number of wave periods spanned by the time range for a given $k_\perp \rho_i$ is found via: 
\begin{equation} 
n = \frac{\Delta t/T_0}{\overline{\omega}_0/\overline{\omega}(k_\perp \rho_i)} ,
\end{equation} 
where $\Delta\ t/T_0$ is the full interval, and $\overline{\omega}(k_\perp \rho_i)$ is determined using the dispersion relation in Appendix A. 

In Fig. \ref{fig:heating}, we plot boxcar averages of the energy injected by the antenna $P_{ant}$ (magenta) and the rate of change of energy contained in the plasma $d \mathcal{W}/dt$ (blue), using an averaging window of $0.5\ t/T_0$. The collisional electron heating $Q_e$ (red) and collisional ion heating $Q_i$ (green) are also plotted, along with the balance between all four components (black). From tracking these quantities, we note that the change in ion energy is negligible, indicating that the electrons are indeed the species of interest as expected for $k_{\perp} \rho_i \gg 1$. Second, we observe that the electrons begin to gain energy significantly after $t/T_0 \approx 0.35$, consistent with the time of saturation of the steady state turbulent energy spectrum. The sum of these rates of change of energy (black) has a systematic offset from zero, which is likely the result of numerical inaccuracies in computing the collisional heating due to the low collisionality chosen in order to avoid smearing of the velocity-space signatures sought in this project. Overall, when considering only the data in the fully saturated interval ($t/T_0 \geq 0.4208$), we find that the average energy conservation rate is $91.53\%$.

\begin{figure}[t]
\includegraphics[width=0.48\textwidth]{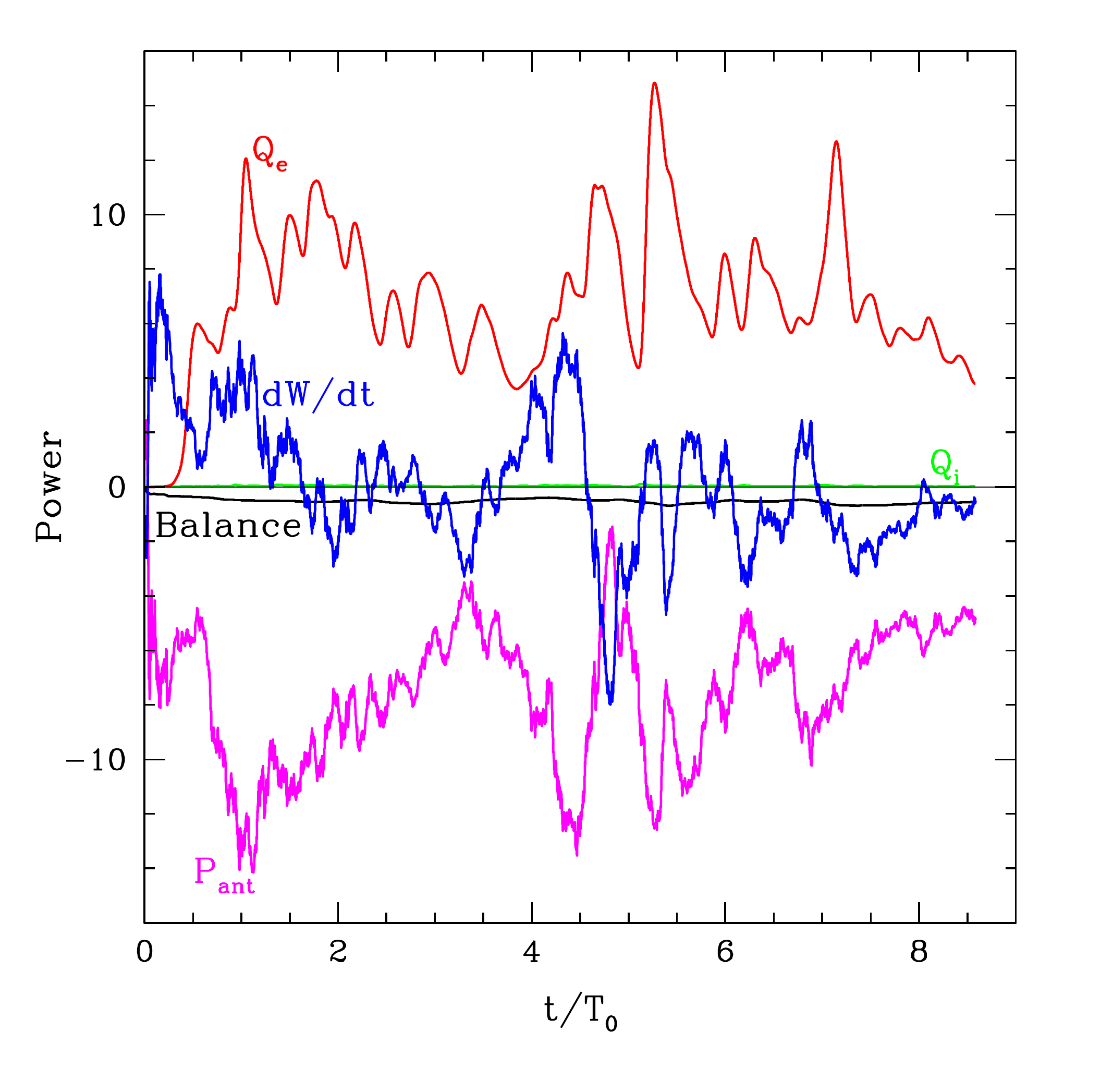}
\caption{\label{fig:heating} Power balance within the \T{AstroGK} \emph{MMS} simulation as a function of time, $t/T_0$. The boxcar average of the antenna and plasma energies $P_{ant}$ and $d \mathcal{W}/dt$ are given, where the boxcar window size is $dt = 0.5\ t/T_0 $. The changes in particle heating are plotted with no averaging ($Q_e$, $Q_i$). The sum of all four values is the energy balance. }%
\end{figure}

\begin{figure}[t]
\includegraphics[width=0.5\textwidth, height=0.5\textwidth]{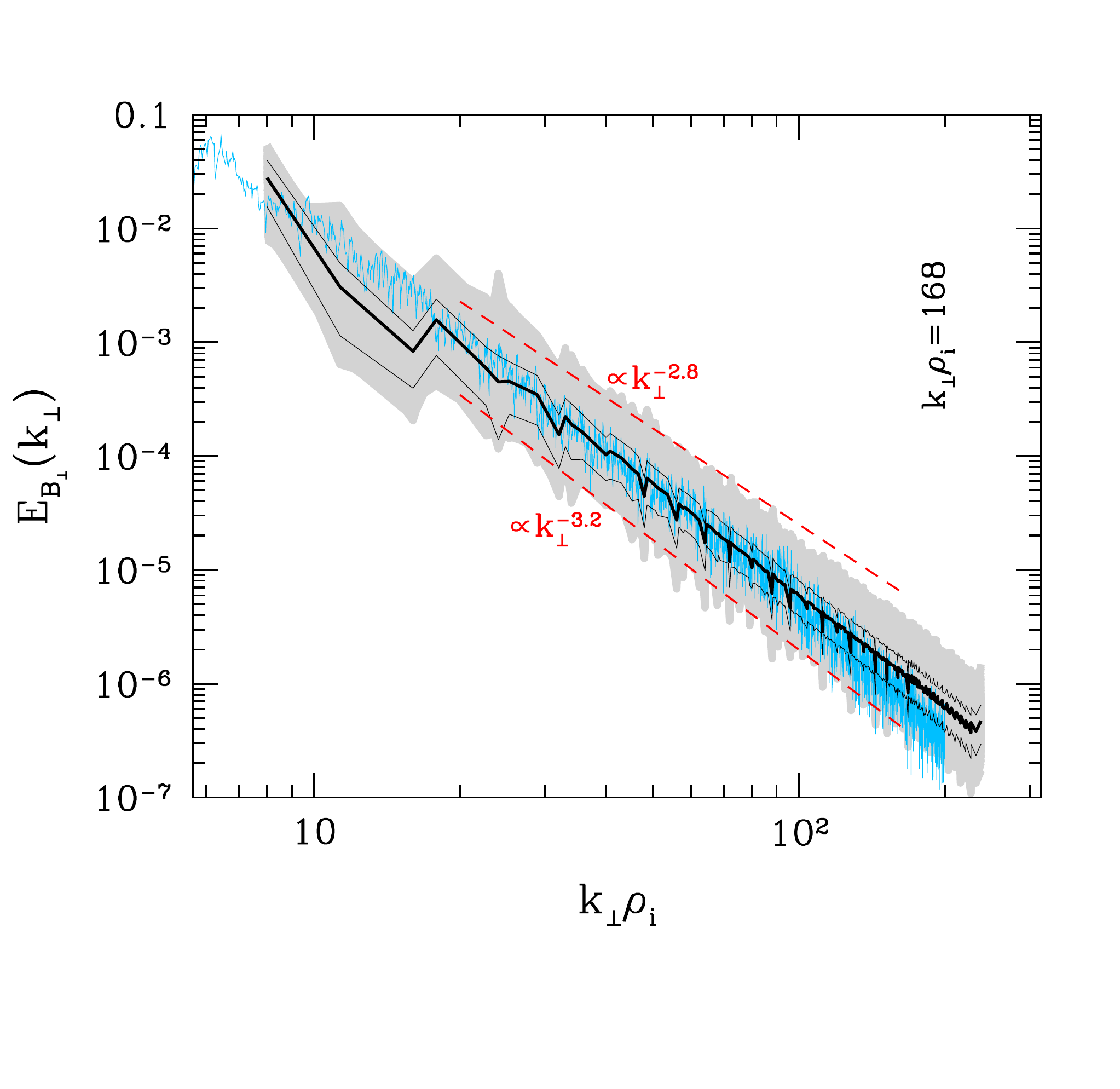}
\vspace{-0.75in}
\caption{\label{fig:E_spec} Magnetic energy spectrum of the \T{AstroGK} \emph{MMS} simulation as a function of perpendicular wavenumber $k_\perp \rho_i$, plotted at all output times after saturation ($t/T_0 \geq 0.4208$) (light gray). The average spectrum (after saturation) is plotted in black, and reference power law slopes in dashed red. We compare the energy spectrum directly to the magnetic energy spectrum from the \emph{MMS} observations (blue).}
\end{figure}

In addition to the energy balance over time, one can also plot the magnetic energy spectrum as a function of perpendicular wavenumber. We obtain a power law spectrum of $E_{B_\perp}(k_\perp)$, shown in Fig.~\ref{fig:E_spec}, as expected for a fully-developed, strong turbulent cascade. Averaging over the full simulation time at steady state, $0.4208 \le t/T_0 \le 8.5834$, we plot the mean turbulent magnetic energy (thick black line) at each resolved value of $k_{\perp} \rho_i$ in the simulation, the standard deviation (thin black lines), and the full range of the instantaneous spectrum over time (light gray). Owing to the nature of the oscillating Langevin input antenna, the spectrum amplitude oscillates over about an order of magnitude, but maintains a strikingly steady shape and slope during these amplitude fluctuations.

We compare the resulting magnetic energy spectrum from the simulation to that from the \emph{MMS} observations (blue) in Fig.~\ref{fig:E_spec}.  We use the Taylor hypothesis \citep{Taylor:1938,Howes:2014a} to convert from a frequency spectrum to a perpendicular wavenumber spectrum using the approximate conversion formula $k_\perp \rho_i \simeq 2 \pi f \rho_i/v_{sw}$, where the $f$ is the linear frequency in Hz, the thermal ion Larmor radius is $\rho_i=57.5$~km, and the magnetosheath plasma flow relative to the spacecraft is $v_{sw}=180$~km/s.  The observations are from the 70~s interval measured by \emph{MMS3} on 16 October 2015 09:24:11--09:25:21; more analysis of this particular interval of magnetosheath turbulence can be found in previous works \citep{Chen:2017,Chen:2019}.
Note that, over the range of wavenumbers, $17 \lesssim k_\perp \rho_i \lesssim 70$, where electron Landau damping is expected to mediate the 
energization of electrons at the expense of turbulent energy (see Appendix~\ref{app:lindisp}), we have excellent agreement of the spectral slope between the simulations and observations.

The upper dashed line marks a power law slope $\propto k_{\perp}^{-2.8}$, which has previously been observed in the dissipation range using both kinetic simulations \citep{Howes:2011,TenBarge:2013a,TenBarge:2013b} and \emph{in situ} observations \citep{Alexandrova:2009,Sahraoui:2010,Alexandrova:2012,Sahraoui:2013b}. 
The spectrum that we find in the dissipation range matches more closely with the lower dashed line, a slightly steeper power law $\propto k_{\perp}^{-3.2}$, which is closer to the spectral slope observed in the \emph{MMS} interval \citep{Chen:2019}. 
In the simulation, the roll off of the spectrum at large $k_{\perp} \rho_i$ indicates that energy is being taken out of the driven system due to a damping mechanism, which produces a steady state spectrum rather than an unphysical energy `pile-up' at the smallest scales. Note that above $k_{\perp}\rho_i = 168$, the steepening of the previously constant slope is a numerical effect rather than a physical one, as we are no longer fully resolving wave vectors at all angles about the magnetic field beyond $k_\perp \rho_i  = 168$.

These analyses give us confidence that there is a well-developed, steady state turbulent cascade in the dissipation range throughout the duration of the simulated \emph{MMS} data, and indicate that energy is indeed being dissipated via some mechanism and secularly transferred to the electrons.

\subsection{Spatial Distribution of Electron Energization}

The above analyses indicate that we will see net energization of electrons within the simulation. As discussed in Sec. \ref{sec:FPC}, this secular energy transfer to the electrons can be traced to the electric field term in the expression for the time rate of change of the phase-space energy density, $w_e({\bf r}, {\bf v}, t)$. For the Landau resonance, this energization is mediated by the parallel electric field interacting with the parallel velocity derivative of the distribution function. Since turbulence is, by nature, intermittent in space and time, we analyze the spatial distribution of the electron energization within the simulation box before turning to the field-particle correlation analysis. 

It can be shown that the change in total kinetic energy of a plasma species $\mathcal{W}_s=\int d^\V{r}\int d^\V{v} m_s v^2 f_s({\bf r}, {\bf v}, t)$ is given by $d\mathcal{W}_s/dt = \int d^3{\bf r}\ {\bf j}_s \cdot {\bf E}$ \citep{Klein:2017, Howes:2017}, where $d\mathcal{W}_s/dt$ is the rate of change of the phase-space energy density for species $s$ (from Sec. \ref{sec:FPC}) integrated over all of phase space. Therefore, by plotting the rate of electromagnetic work, ${\bf j}_s \cdot {\bf E}$, across the simulation space, we indicate the locations where net energy exchange is likely to be occurring, and whether the overall transfer is net positive or negative in the simulation domain. Note that for the anisotropic \Alfvenic turbulence modeled in this simulation, the total ${\bf j}_s \cdot {\bf E}$ is dominated by its parallel contribution $j_{\parallel s} E_\parallel$. Furthermore, at the small scales $k_\perp \rho_i \gg 1$ modeled in this simulation, the parallel current is dominated by the electron motion, $j_\parallel \simeq j_{\parallel e}$, so we focus here on plots of $j_{\parallel}$, $E_\parallel$, and their product over the simulation domain.

\begin{figure}[t]
\includegraphics[width=0.5\textwidth, height=0.4\textwidth]{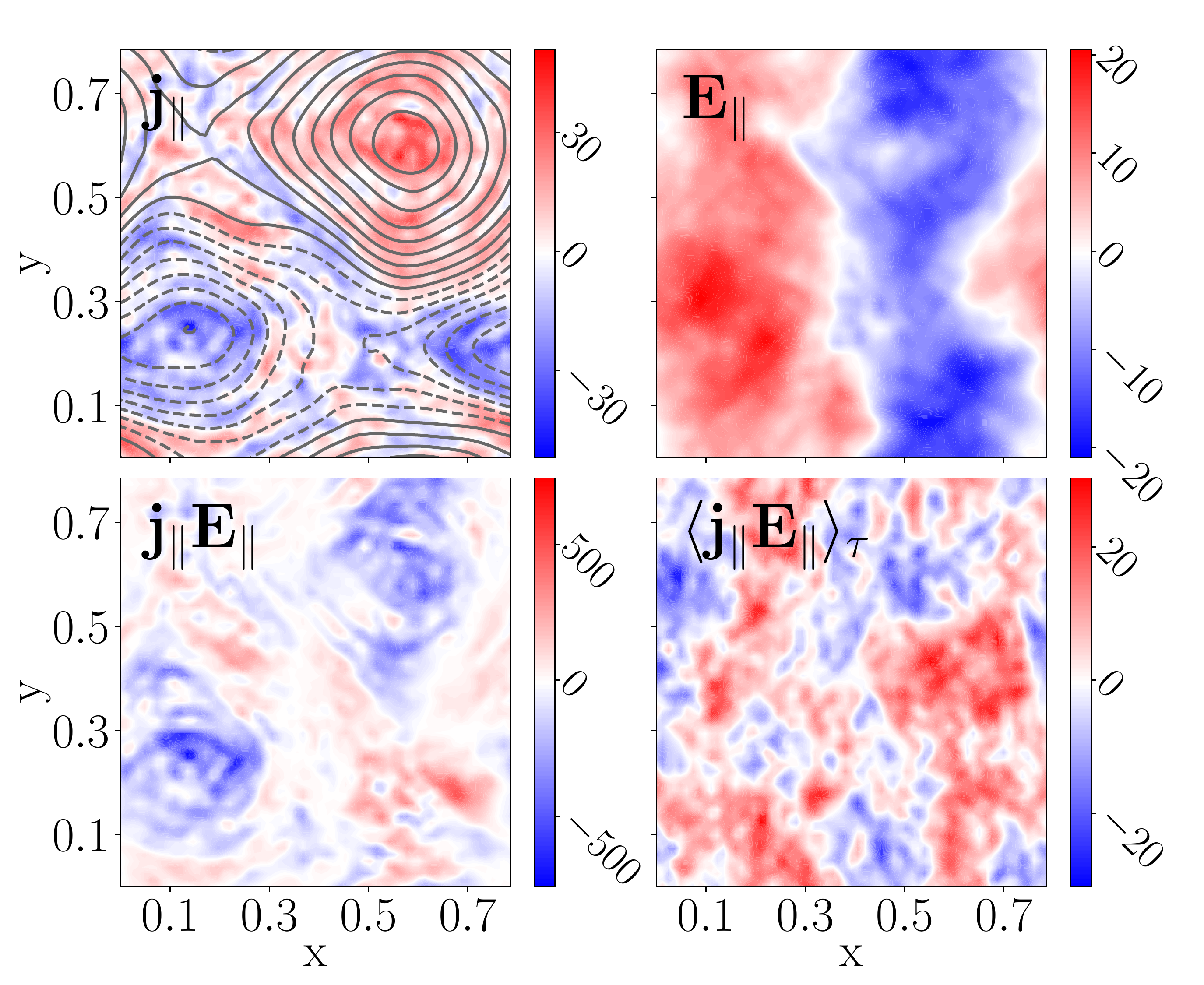}%
\caption{\label{fig:jdotE}  Simulation data in a cross-section of the simulation box at $z=0$ and time $t/T_0 =2.53$: the parallel current ($j_\parallel$), overlaid with contours of $A_\parallel$ (solid: +ve; dashed: -ve); the parallel electric field $ E_\parallel$; the instantaneous electromagnetic work $j_{\parallel}$, $E_\parallel$, and the electromagnetic work time-averaged over four outer-scale Alfv\'en wave periods $\tau/T_0 = 4$ centered at time $t/T_0 =2.53$, $\langle j_\parallel  E_\parallel \rangle_\tau$.}
\end{figure}

Each panel of Fig. \ref{fig:jdotE} is a cross-section of the $x$-$y$ plane of the simulation box at $z=0$, and the data (in the first three panels) are taken from the same simulation output time, $t/T_0 = 2.53$. The upper left panel, $j_\parallel$, shows the current parallel to the mean magnetic field (${\bf B}_0 $), overlaid with contours of the parallel vector potential. Though some small-scale structures are visible in $j_\parallel$, the snapshot of parallel current and vector potential is dominated by structures on the order of the simulation size, $L_\perp$, which corresponds to Alfv\'en waves at the driving scale, $k_{\perp 0} = 2\pi/L_\perp$. The upper right panel shows a snapshot of the parallel electric field, $E_\parallel$, where the structures are again on the order of the driving mode.

In the two bottom panels, we plot the parallel contribution to the electromagnetic work. On the left is a snapshot of $j_\parallel  E_\parallel$ at $t/T_0 = 2.53$, which is also dominated by driving-scale structures that have little effect on the secular electron energization. After correlating over a sufficiently long time interval, however, the large scale wave modes average out to reveal small amplitude structure in the higher-order modes of the dissipation range. This is shown on the right, where we plot $\langle j_\parallel  E_\parallel \rangle_\tau$, averaged over a time interval $\tau/T_0 = 4$
centered on the same snapshot time as the other three panels. Note that in  $\langle j_\parallel  E_\parallel \rangle_\tau$, the dominant structures have indeed become much smaller than the driving scale, and that the amplitudes of the snapshot and time-averaged electromagnetic work differ by a factor of $30$. This large discrepancy in amplitudes is what we would expect if the net energization is dominated by scales much smaller than the driving scale of the turbulence and there is a steep magnetic energy spectrum, such as that shown in Fig. \ref{fig:E_spec}. 

When the instantaneous current, electric field, and electromagnetic work in the first three panels are plotted over the duration of the simulation, we observe that they oscillate in time. By comparison, however, the correlated electromagnetic work shows marked temporal consistency, indicating that at the smallest scales, the particles in the simulation are able to experience secular, rather than oscillatory, transfers of energy. Additionally, we note that $\langle j_\parallel  E_\parallel \rangle_\tau$ is largely positive across the $z=0$ plane throughout the simulation, indicating net positive work being done by the fields on the electrons, on average over the spatial domain.

\begin{figure*}[t]
\hbox{
\hfill
\vspace{-0.19cm}
\resizebox{0.5\textwidth}{!}{\includegraphics[width=0.5\textwidth]{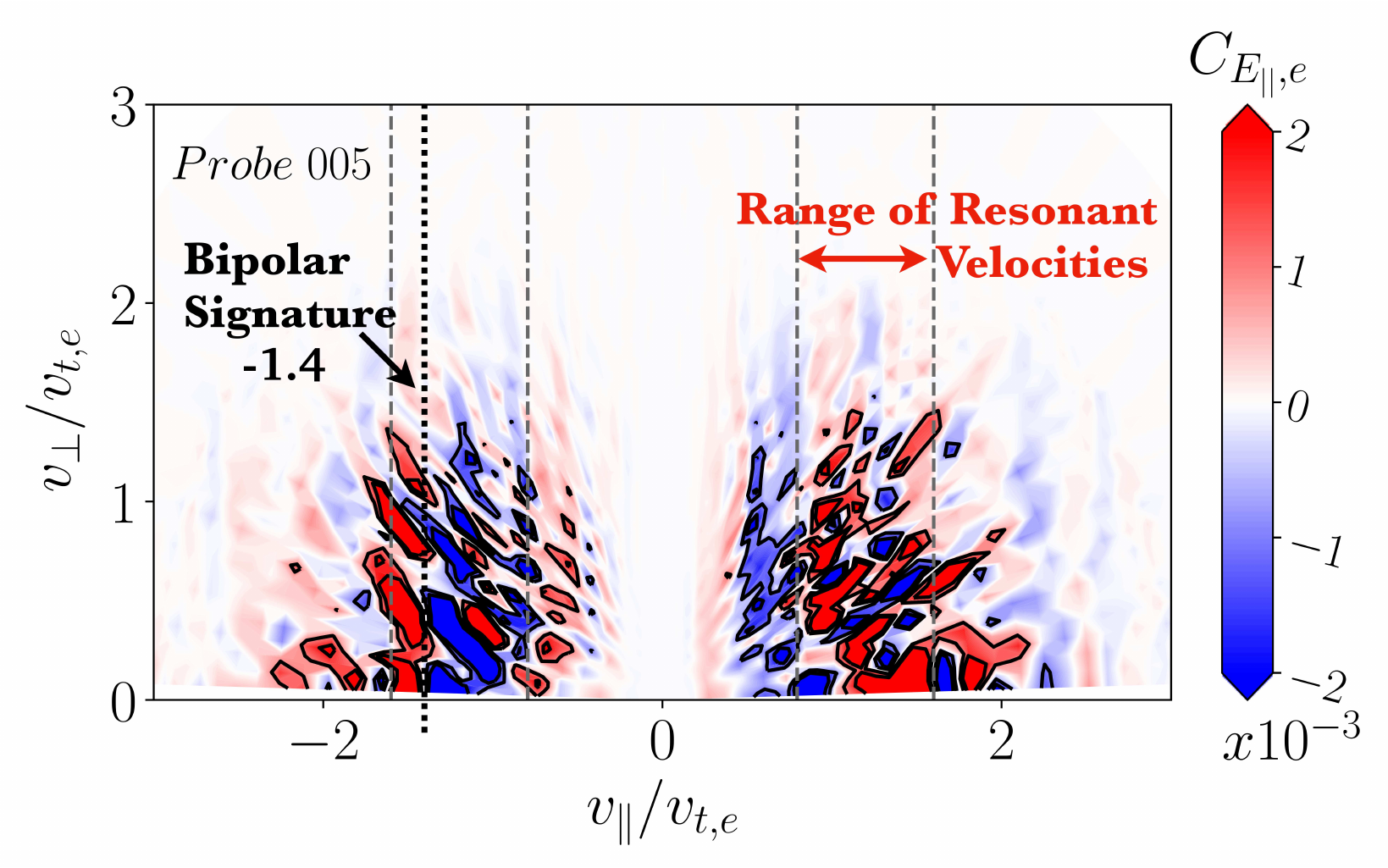}}
\resizebox{0.5\textwidth}{!}{\includegraphics[width=0.5\textwidth]{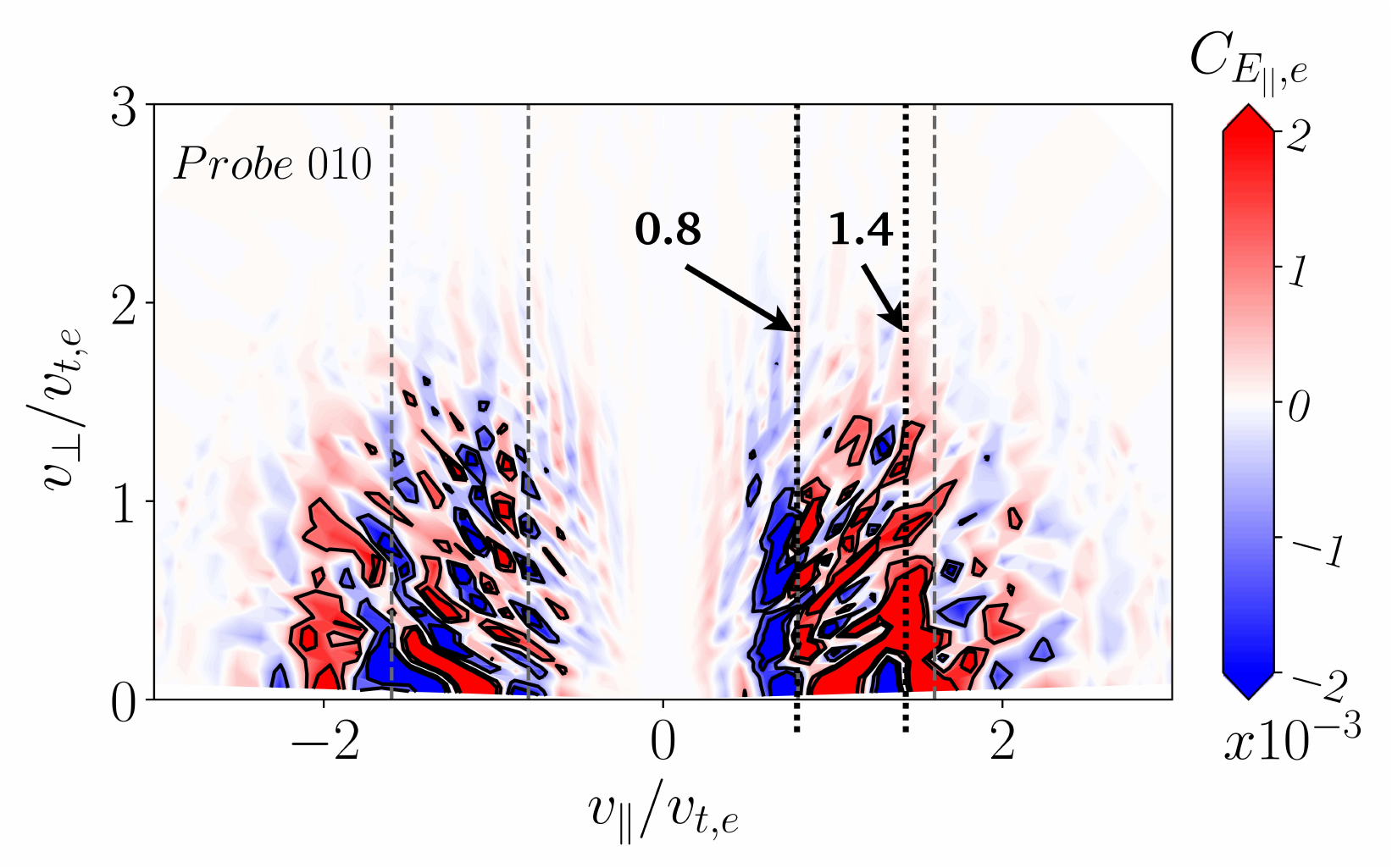}}}
\hbox{
\hfill
\resizebox{0.5\textwidth}{!}{\includegraphics[width=0.5\textwidth]{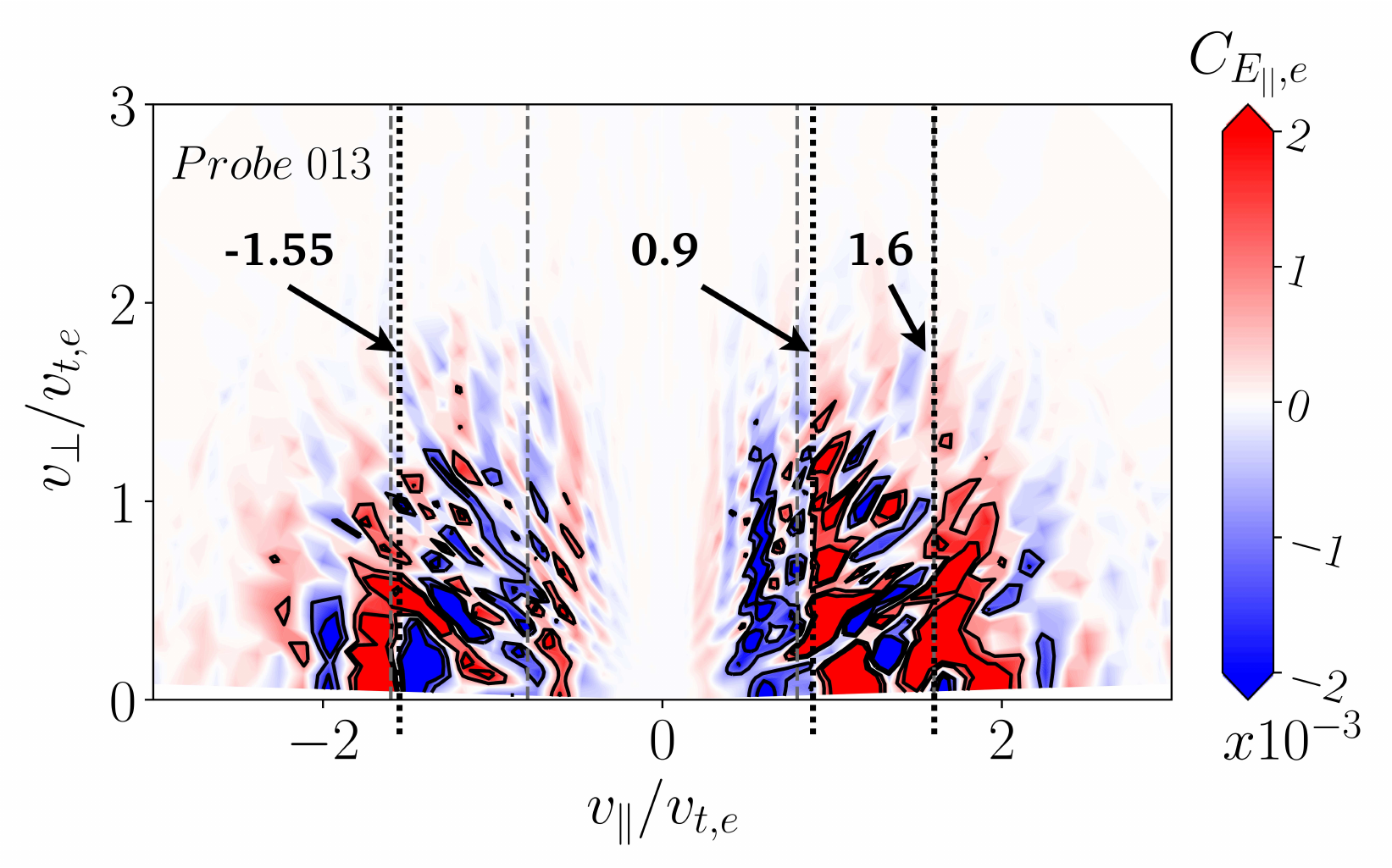}}
\resizebox{0.5\textwidth}{!}{\includegraphics[width=0.5\textwidth]{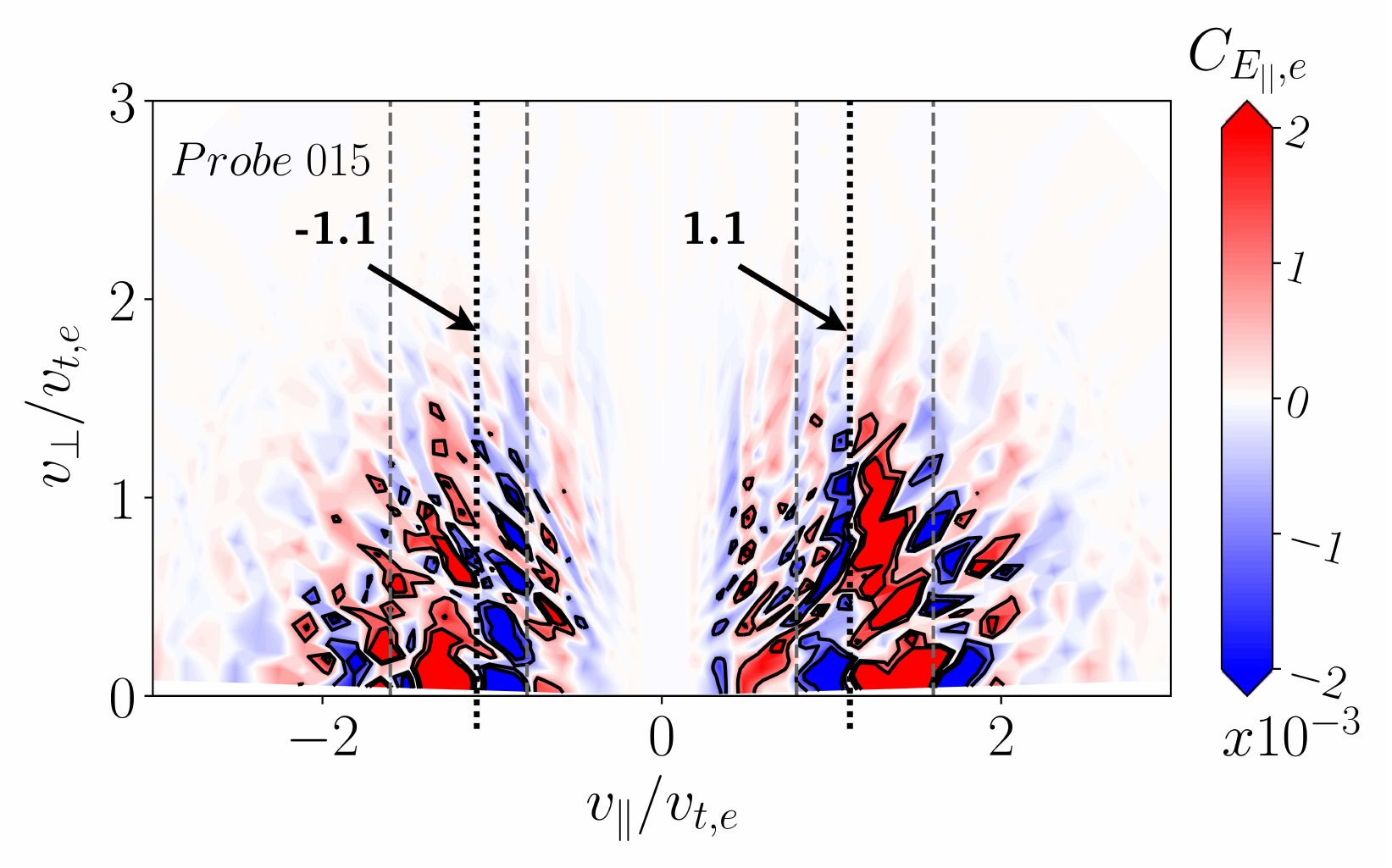}}} 
\caption{ \label{fig:gyroplots}  The gyrotropic field-particle correlation $C_{E_\parallel, e}(v_{\parallel},v_{\perp})$ for electrons at selected \emph{MMS} simulation probes using a correlation interval of $\tau/T_0 = 4$ centered at time $t/T_0=2.53$. Dashed lines represent the range of resonant velocities expected to be important for Landau damping of dispersive kinetic Alfv\'en waves in the dissipation range: $0.79 \le |v_{\parallel}/v_{t,e}| \le 1.6$. Bipolar signatures indicating electron Landau damping at $v_\parallel/v_{t,e}=\omega/k_{\parallel} v_{t,e}$ are marked with dotted lines and arrows.}
\end{figure*}

\section{\label{sec:FPCA} Field-Particle Correlation Analysis}
\subsection{Gyrotropic Velocity-Space Signatures}

Next, we turn to a field-particle correlation analysis of the electron energization, hoping to uncover similarities to our motivating observational \emph{MMS} interval and to gain new insight into electron Landau damping in the heliosphere. All twenty-four probe points in the simulation box were analyzed using this technique, and for the purposes of this work we select probes 5, 10, 13, and 15 from the $z=0$ plane as examples of points where the electromagnetic work by $E_\parallel$ on the electrons is largely positive throughout the duration of the interval. This choice is motivated by the findings in Sec. \ref{sec:AGKS}, where the time-averaged electromagnetic work $\langle j_\parallel  E_\parallel \rangle_\tau$ was found to be largely positive in the $z=0$ plane. We apply the field-particle correlation method to the single-point data of $E_{\parallel}$ and $\partial f_e/ \partial v_\parallel$ at the probes, as described in Sec. \ref{sec:FPC}, again choosing an interval  corresponding to four outer-scale Alfv\'en wave periods. The resulting correlation, $C_{E_{\parallel}}({\bf v}, t/T_0)$, is essentially a sliding time-average over the correlation interval, $\tau/T_0=4$, which we plot in gyrotropic velocity space ($v_{\parallel}$, $v_{\perp}$) for a snapshot in time. In Fig. \ref{fig:gyroplots}, we plot the gyrotropic correlation for our selected probes at $t/T_0 = 2.53$, where we label the time of the snapshot by the time at the center of the correlation interval.

\begin{figure*}[t]
\hbox{
\hfill
\vspace{-0.19cm}
\resizebox{0.5\textwidth}{!}{\includegraphics[width=0.5\textwidth]{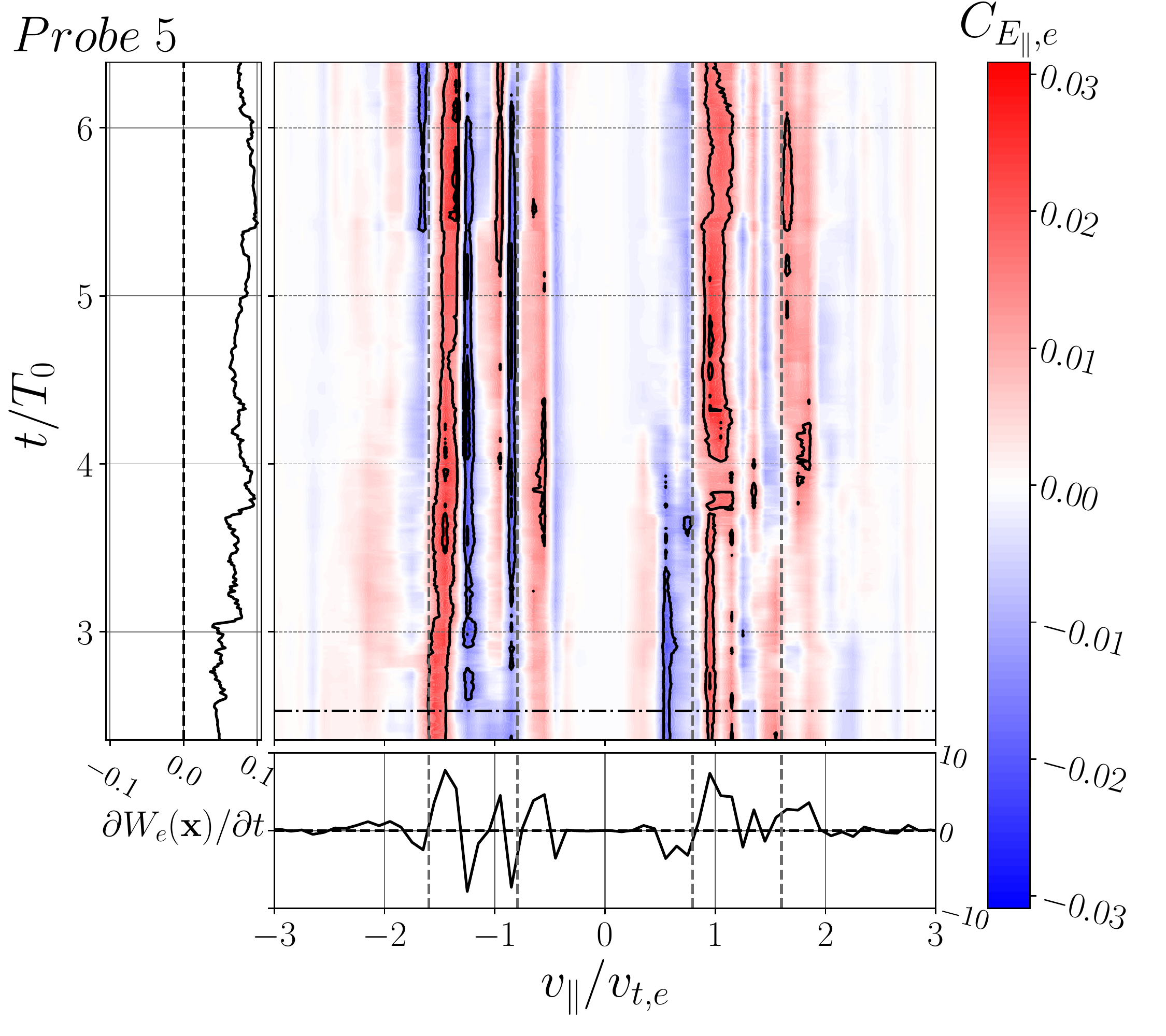}}
\resizebox{0.5\textwidth}{!}{\includegraphics[width=0.5\textwidth]{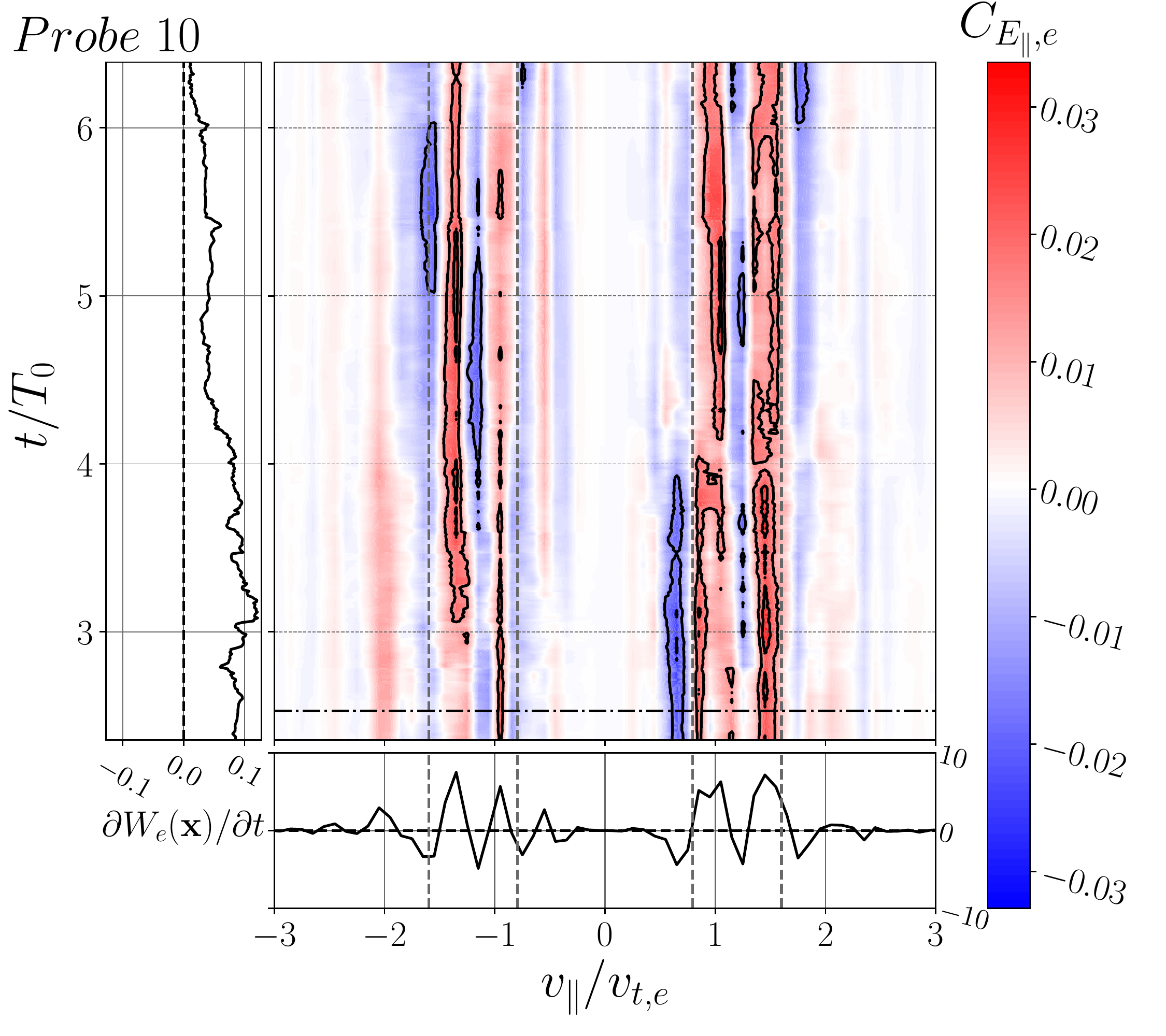}}}
\hbox{
\hfill
\resizebox{0.5\textwidth}{!}{\includegraphics[width=0.5\textwidth]{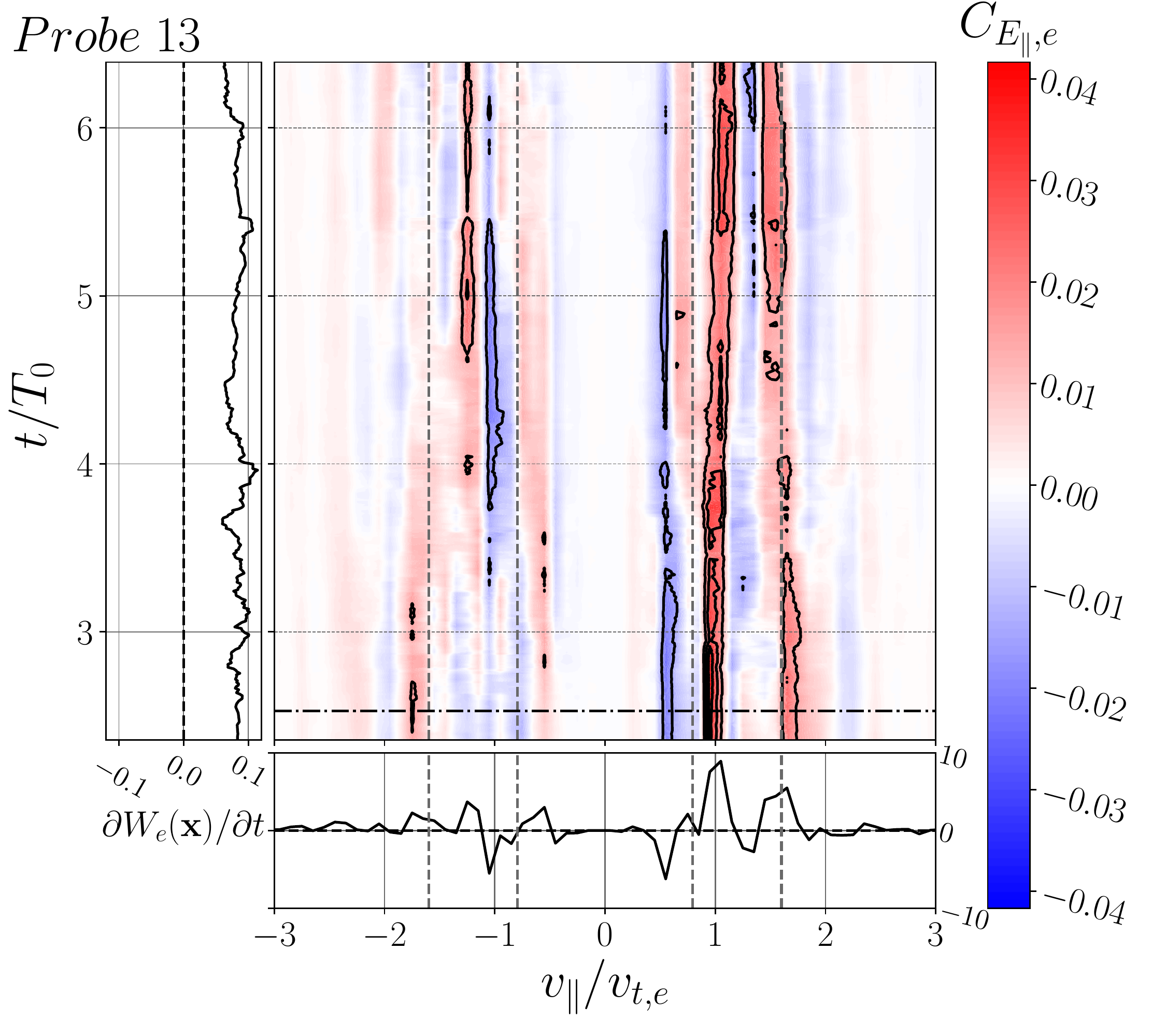}}
\resizebox{0.5\textwidth}{!}{\includegraphics[width=0.5\textwidth]{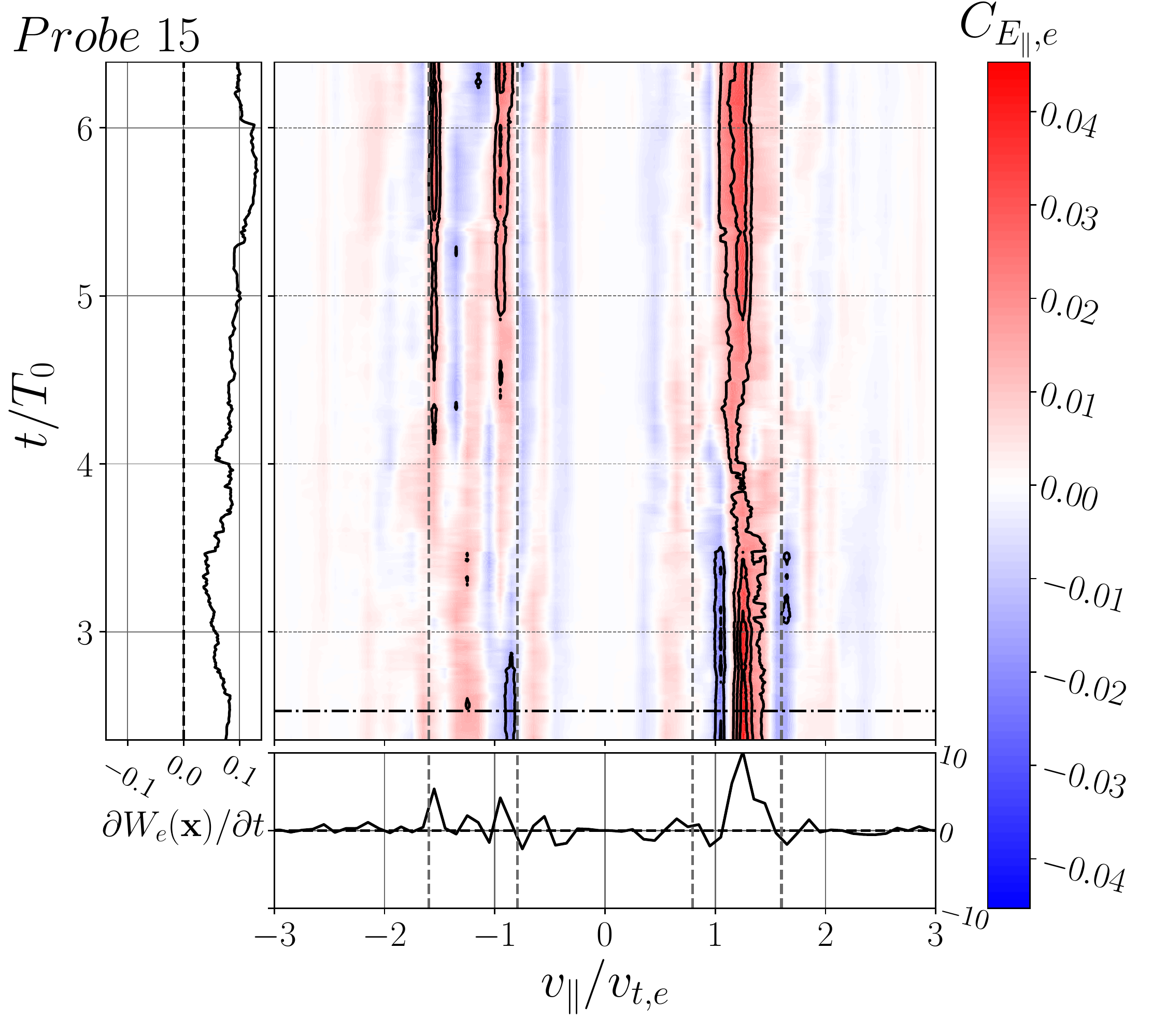}}}
\caption{ \label{fig:heatstacks} Timestack plots of the reduced parallel field-particle correlation $C_{E_\parallel, e}(v_{\parallel},t)$ for electrons at selected \emph{MMS} simulation probe points in the $z=0$ plane, where vertical dashed lines show the expected range of parallel resonant velocities $v_\parallel/v_{t,e}=\omega/k_{\parallel} v_{t,e}$. Note the persistence of many resonant signatures across long stretches of simulation time. The left subplots present the persistently positive rate of change of spatial energy density found by integrating over parallel velocity, $\partial W_e/\partial t = \int dv_\parallel C_{E_\parallel, e}(v_{\parallel},t)$, and the lower subplots present the time-integrated parallel velocity-space signature,  $C_{E_\parallel, e}(v_{\parallel}) = \int dt C_{E_\parallel, e}(v_{\parallel},t)$.}
\end{figure*}

In these gyrotropic plots, at various points along the $v_{\parallel}$ axis we see the bipolar heating signatures characteristic of Landau damping (Sec. \ref{sec:PEDT}). Ion Landau damping results in gyrotropic velocity-space plots dominated by a single bipolar signature \citep{Klein:2017} due to the relatively nondispersive nature of KAWs at the ion scales (as discussed in Sec.~\ref{sec:diagram} and seen in the linear dispersion relation plot in Fig.~\ref{fig:LGKDR}), but Fig. \ref{fig:gyroplots} shows energization patterns that are  more `cluttered.' There is a combination of apparently unpatterned energization along with the typical bipolar signatures at different parallel velocities. Often, we observe multiple bipolar signatures in a single gyrotropic plot, consistent with our expectations for simultaneous damping of various dispersive KAW modes with different values of $k_\perp \rho_i$ and therefore different parallel phase velocities $\omega/k_\parallel$. 

In Fig. \ref{fig:gyroplots}, the vertical dashed lines on top of the correlation $C_{E_\parallel, e}(v_{\parallel},v_{\perp})$ indicate the range of resonant parallel velocities we expect to be significant in electron Landau damping, $0.79 \le |v_{\parallel}/v_{t,e}| \le 1.6$. In each of the four plots, the most prominent bipolar signatures are indicated by arrows and a vertical dotted line through the resonant velocity, labeled with the magnitude of the parallel electron velocity $v_\parallel/v_{t,e}=\omega/k_{\parallel} v_{t,e}$. For probe 5, the prominent signature is at $v_{\parallel}/v_{t,e} = -1.4$, indicating damping of a KAW propagating down the magnetic field. For probe 10, we mark prominent signatures from two different wavemodes propagating up the magnetic field, at $v_{\parallel}/v_{t,e} =0.8$ and $1.4$. We see bipolar signatures for both positive and negative wave modes in probe 13, at $v_{\parallel}/v_{t,e} =-1.55$, $0.9$, and $1.6$. This is also observed for probe 15, where the damping occurs symmetrically at $v_\parallel/v_{t,e} = \pm 1.1$.

\subsection{Behavior of the Velocity-Space Signatures in Time}

The gyrotropic plots of $C_{E_\parallel, e}(v_{\parallel},v_{\perp})$ show the rate of change of phase-space energy density at a single point in time (averaged over the correlation interval $\tau$). Knowing the intermittent nature of turbulence in time and space \citep{Zhdankin:2015a,Zhdankin:2015b,Zhdankin:2016,Mallet:2019}, we must then ask whether these energization signatures are persistent in time. To answer this, we integrate the gyrotropic field-particle correlation  over $v_{\perp}$ to obtain the reduced parallel field-particle correlation at each point in time,  $C_{E_{\parallel, e}}( v_\parallel,t)\int dv_\perp C_{E_\parallel,e} (v_{\parallel},v_{\perp},t)$. Note that the reduced parallel field-particle correlations $C_{E_{\parallel, e}}( v_\parallel)$ are computed at each centered time $t$ as our correlation interval $\tau$ slides over the full simulation duration, and the resulting one-dimensional correlations are `stacked' in time to construct \textit{timestack plots} of $C_{E_{\parallel, e}}( v_\parallel,t)$, shown in Fig.~\ref{fig:heatstacks}. 

Each of the four panels in Fig.~\ref{fig:heatstacks} presents a composite plot: the center shows a contour plot of the timestack $C_{E_{\parallel, e}}( v_\parallel,t)$; the left subplot presents the rate of change of spatial energy density over time by integrating over parallel velocity, $\partial W_e/\partial t = \int dv_\parallel C_{E_\parallel, e}(v_{\parallel},t)$; and the lower subplot presents the time-integrated parallel velocity-space signature,  $C_{E_\parallel, e}(v_{\parallel}) = \int dt C_{E_\parallel, e}(v_{\parallel},t)$. These time-integrated parallel velocity-space signatures may obscure some relatively small magnitude, short duration energizations, but they nicely give an overall picture of the locations in $v_{\parallel}/v_{t,e}$ where the most significant resonances are located. The locations of the bipolar resonances in the contour plot appear clearly as zero-crossings in the time-integrated plots, where positive slope for $v_{\parallel}/v_{t,e} >0$, or negative slopes for $v_{\parallel}/v_{t,e}  < 0$, are characteristic signatures of Landau damping. 

The velocity-integrated plots, as explained in Sec.~\ref{sec:PEDT}, yield the change in total spatial energy density of the electrons, $dW_e/dt$. As expected, for each the four highlighted probes, this change in the electron energy is consistently positive throughout the simulation, indicating transfer of net energy to the electrons. It is important to note that $dW_e/dt$ was not consistently positive for all probe points; however, for the majority (thirteen out of twenty-four probes) the total change in energy was positive throughout the entire simulation interval, and it was consistently negative for none. 

An interesting feature brought to light in the timestack plots is the appearance of bipolar signatures of electron Landau damping at different parallel velocities, $v_\parallel/v_{t,e}$, apparently corresponding to KAWs with different values of $k_\perp \rho_i$, yielding different parallel phase velocities, $\omega/k_\parallel v_{te}$, due to the dispersive nature of KAWs in this regime. Some signatures appear to fade in and out of significance with time, and others that are nearby in velocity-space seem to couple together to form a superposed signature in the time-integrated parallel correlation $C_{E_\parallel, e}(v_{\parallel})$, as in the theoretical example discussed in Sec.~\ref{sec:diagram}. Below, we point out some of the most prominent features that we observe for these four example probes.

The time of the gyrotropic plots in Fig.~\ref{fig:gyroplots}, at $t/T_0 = 2.53$, is marked in Fig.~\ref{fig:heatstacks} with a horizontal dashed line across the contour plots. If we compare this line in the timestack plot of probe 5 with its gyrotropic counterpart, we find that the resonance at  $v_{\parallel}/v_{t,e} = -1.4$ is clearly visible in both plot types, and from the timestack plot we see that it persists until the end of the simulation. Additionally, the `cluttered' energization at $ v_\parallel > 0$ in the gyrotropic plot has been condensed by the $v_\perp$-integration into a faint signature roughly centered on the lower bound of the range of significant resonances at $t/T_0 = 2.53$. This could be an artifact of how the disordered energization pattern at $v_\parallel >0$ in the gyrotropic plane happened to sum together under the $v_\perp$-integration; however, since the timestack shows that the region of phase-space just above $v_\parallel = 0.79$ experiences persistent heating throughout the simulation, it seems credible that at $t/T_0 = 2.53$, a small-amplitude signature has begun to develop at $v_\parallel/v_{t,e} = 0.79$, but is obscured at that time in the gyrotropic plane either by its faintness or other overlapping resonances. These two resonances ($v_\parallel/v_{t,e} = -1.4$ and $0.79$) are also visible as zero-crossings in the time-integrated correlation of probe 5, along with a weaker resonant signature around $v_\parallel/v_{t,e} =-0.9$, and another outside the expected range at roughly $v_\parallel/v_{t,e} =-0.5$.

For probe 10, we see that both of the positive parallel velocity signatures that are visible in the gyrotropic snapshot ($v_\parallel/v_{t,e} = 0.8$ and $1.4$) are persistent throughout the simulation. The timestack plot also indicates that a faint signature may exist at $v_\parallel/v_{t,e} = -0.8$ at the time center of $t/T_0 = 2.53$, though it is not easily identifiable in the gyrotropic plot. Also note that after about $t/T_0 = 3.1$, we see another signature develop around $v_\parallel/v_{t,e} = -1.2$. 

In the plots of probe 13, the resonance at $v_\parallel/v_{t,e} = 1.6$ that was prominent in the gyrotropic plot is observed to fade in significance with time, and is replaced by a signature at $v_\parallel/v_{t,e} = 1.4$. It is interesting to note that, in the time-integrated correlation for probe $13$, these two signatures blend into a single broadened peak, with a zero-crossing at $\sim1.4$. This example indicates how kinetic Alfv\'en waves in the dispersive range may have overlapping velocity-space signatures that lead to a broader combined signature, as was theoretically illustrated in Fig.~\ref{fig:LDdiagramFull}. The resonance at $v_\parallel/v_{t,e} = 0.9$ seen in the gyrotropic plot also appears to `drift' to higher velocities in the timestack, possibly as damping of a higher phase-velocity Alfv\'en wave becomes dominant. The signature we observed at $v_\parallel/v_{t,e} = -1.55$ quickly fades with time. 

Finally, the timestack plot of probe 15 shows that the set of symmetric signatures visible in the gyrotropic plot, at $v_\parallel/v_{t,e} =\pm 1.1$, have very different behaviors with time. The energization signature at the negative phase-velocity resonance is faint and quickly fades as the simulation progresses, while the signature at the positive phase-velocity resonance persists throughout the simulation. 

Together, all of these results indicate the much more complicated behavior of electron Landau damping compared to the behavior of ion Landau damping, due to the energy transfer being dominated by KAWs with different values of $k_\perp \rho_i$, and thus different parallel resonant phase velocities $\omega/k_\parallel$. The resulting velocity-space signatures of the electron Landau damping are consequently a complicated superposition of these different resonant energy transfers, yielding reduced parallel correlations of widely varying qualitative appearance from case to case, as seen in the four time-integrated plots of $C_{E_\parallel, e}(v_{\parallel})$ shown in Fig.~\ref{fig:heatstacks}.

\subsection{Long-Time Average of $C_{E_\parallel, e}(v_{\parallel})$}

One advantage of numerical data over spacecraft data is the ability to probe multiple single-point locations with one simulation. However, a distinct disadvantage is the limited feasible duration of the simulation due to computational cost. Observational data typically represent a single spatial point per instrument, but have an abundance of time. This \emph{MMS} simulation, though long duration for a high-resolution synthetic plasma, only spans $8.16$ driving-scale kinetic Alfv\'en wave periods at wavenumber $k_{\perp 0}\rho_i = 8$. The observational \emph{MMS} interval used in CKH19, by contrast, encompasses $280$ Alfv\'en cycles of the same frequency. This is found by assuming the validity of the Taylor hypothesis, which yields that wave modes at $k_{\perp 0} \rho_i = 8 $ in the simulation correspond to waves at a frequency of about 4~Hz in the spacecraft frame \citep{Taylor:1938,Howes:2014a,Chen:2019}.

\begin{figure}[t]
\includegraphics[width=0.45\textwidth, height=0.4\textwidth]{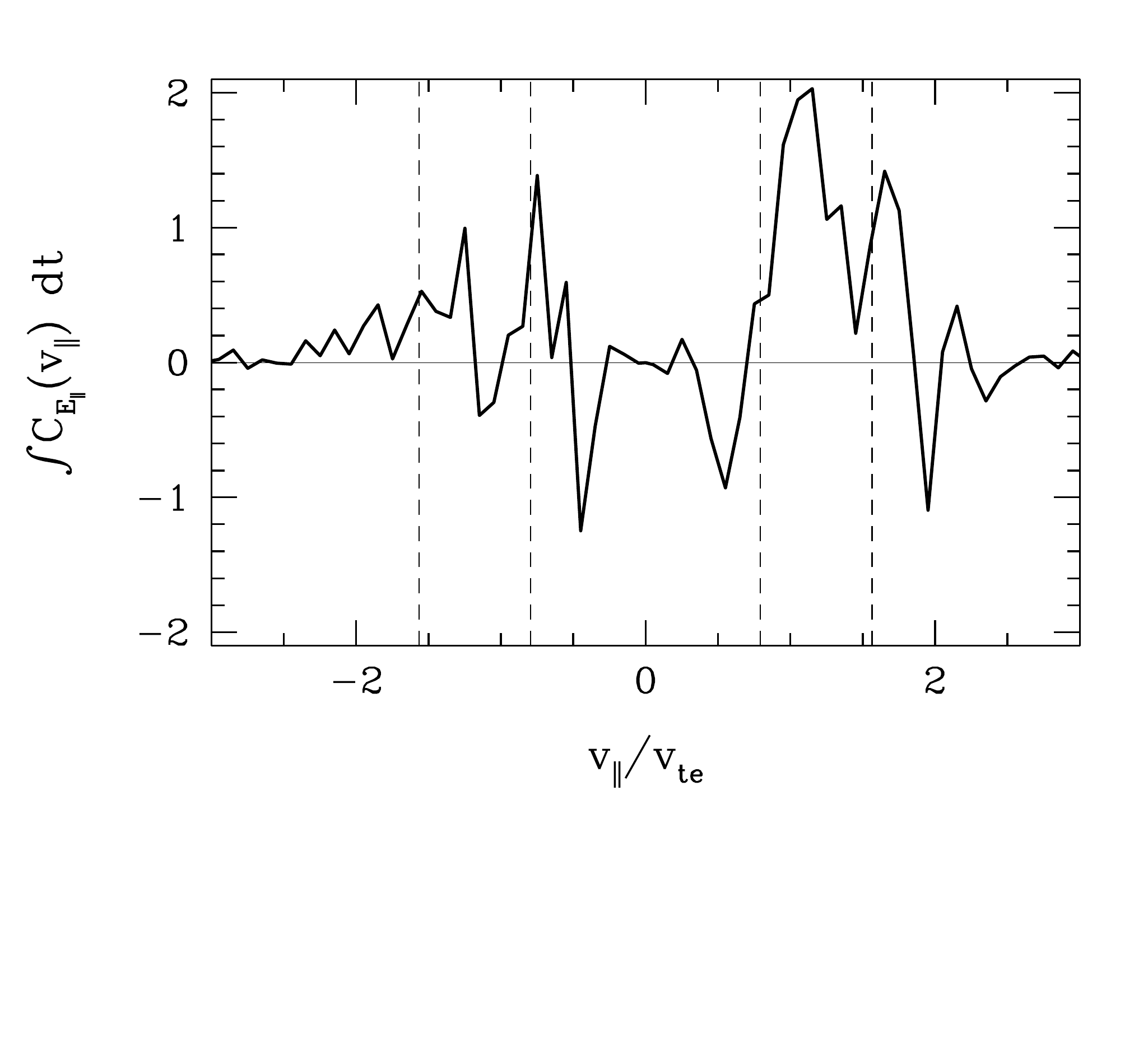}
\vspace{-0.8in}
\caption{\label{fig:avg_intC} Average of the time-integrated $C_{E_{\parallel,e}}(v_\parallel)$ correlations from all twenty-four simulation probe points, where the effective time-integration is over an interval spanning $196$ KAW modes at $k_{\perp 0} \rho_i = 8$. Vertical dashed lines indicate the expected range of resonant parallel velocities, $0.79 \le |v_{\parallel}/v_{t,e}| \le 1.6$.}%
\end{figure}

In CKH19, the authors find that when the field-particle correlated data are time-averaged over the full 70~s interval (similar to the process used to generate the lower panels in Fig.~\ref{fig:heatstacks}), a signature consistent with Landau damping emerges with two clear zero-crossings near the electron thermal speed for positive and negative propagating waves \citep{Chen:2019}. A qualitative comparison between the CKH19 result and a similar result from this simulation would be useful, yet the length of time over which any of the simulated single-point correlations can be integrated differs from the CKH19 time interval by a factor of $\sim35$. However, we can utilize the multiple single-point locations to our advantage and effectively extend our time interval. To do this, we assume that our probe points are sufficiently removed from each other in physical space so as to make the turbulent fields at each point essentially independent. Such an assumption is supported by the intermittent nature of turbulent dissipation, where the rate of plasma energization varies on small scales in the directions perpendicular to the local mean magnetic field \citep{Zhdankin:2014,Howes:2018}. Under this assumption, we average the twenty-four individual time-integrated correlations to produce a single plot of an effective simulation extending over approximately $8.16 \times 24 = 196$ outer-scale Alfv\'en wave periods, reducing the time discrepancy between the simulated and observational intervals from a factor of $\sim35$ to a factor of $\sim1.4$. The result of this integration is shown in Fig.~\ref{fig:avg_intC}. In the next section, we discuss the direct comparison of our numerical results to the \emph{MMS} observational results\cite{Chen:2019}. 

\section{\label{sec:D} Discussion}
\subsection{Gyrotropic Velocity-Space Signatures}

Figure 1(c) of CKH19 presents a gyrotropic plot of the alternative field-particle correlation $C'_{E_\parallel, e}(v_\parallel,v_\perp) = q_e v_\parallel f_e(v_\parallel,v_\perp) E_\parallel$.  The relation between the standard correlation used in this study, $C_{E_\parallel, e}$, and the alternative correlation, $C'_{E_\parallel, e}$, presented there is given by:
\begin{equation}
C_{E_\parallel, e}(v_\parallel,v_\perp) = -\frac{v_\parallel}{2} \frac{\partial C'_{E_\parallel, e}(v_\parallel,v_\perp)}{\partial v_\parallel} + \frac{C'_{E_\parallel, e}(v_\parallel,v_\perp)}{2} .
\end{equation}
In practice, the zero-crossings in $v_\parallel$ of the standard correlation correspond roughly to maxima or minima in $v_\parallel$ in the alternative correlation.  Specifically, the positive peaks in $v_\parallel$ of $C'_{E_\parallel, e}$ would correspond to the bipolar zero-crossings from negative to positive in $C_{E_\parallel, e}$ that indicate energization of electrons by Landau damping.

Qualitatively, the gyrotropic plot of $C'_{E_\parallel, e}(v_\parallel,v_\perp)$ shown in Figure 1(c) of CKH19 shows clear signatures (peaks in $C'_{E_\parallel, e}$) indicative of Landau damping, with two symmetrically-located peaks just above the positive and negative electron thermal velocity. Overall, the velocity-space signature of $C'_{E_\parallel, e}$ is very smooth in appearance in the observations, with the corresponding symmetric zero-crossings in the standard correlation  $C_{E_\parallel, e}$ at $v_\parallel/v_{t,e}=\pm 1.2$, as shown in Fig 2(c) of CKH19. The gyrotropic plots of  Fig.~\ref{fig:gyroplots} from the simulation indeed show bipolar signatures in $C_{E_\parallel, e}(v_\parallel,v_\perp)$, indicated by the vertical dotted lines, but their qualitative appearance and quantitative location vary widely from point to point. In addition to the clear bipolar signatures visible  within the range of resonant parallel velocities, $0.79 \le |v_{\parallel}/v_{t,e}| \le 1.6$, also seen in our simulation results is a variety of fainter `clutter': perhaps comprised of the signatures of faint Landau damping either just developing or fading out, or of small-scale oscillatory energization that was not averaged out over the correlation interval $\tau$ along with the large-scale oscillations. 

Several factors may contribute to the salient qualitative differences between the results of the CKH19 observation and our simulation. First, although the \emph{MMS} spacecraft velocity distribution measurements have the highest resolution of any operating mission, the velocity-space of our simulation has even higher resolution. The simulation has a polar grid of $n_\varepsilon \times n_\lambda = 32 \times 128=4096$ velocity grid points in 2V gyrotropic velocity space $(v_\parallel,v_\perp)$ over the range $-3 \le v_\parallel/v_{t,e} \le 3$ and $0 \le v_\perp/v_{t,e} \le 3$.  The electron measurements of the Dual Electron Spectrometer (DES) instrument on \emph{MMS} employ $n_E \times n_\theta \times n_\phi = 32 \times 16 \times 32 = 16834$ velocity-space measurements in full 3V velocity space \citep{Pollock:2016} over a possibly much larger range of velocities, minimally $0.5 \le v/v_{t,e} \le 6$.  A rough estimate, accounting for the different dimensionalities of these grids, is that the simulation velocity-space resolution is at least a factor of 4 better than the resolution of the observation. If features exist in each dataset between this gap in resolution, they would appear distinctly in the simulated velocity-space but could lead to some averaging in the correlation of the observational result, potentially yielding a single, broad signature without small-scale features. Future work will attempt to quantify how the difference in velocity-space resolution leads to qualitative differences between the observational and numerical velocity-space signatures of the electron energization. 

Another difference could be due to smearing of the observational signature in $(v_\parallel,v_\perp)$ space due to fluctuations in the direction of the magnetic field over the measurement interval. The \emph{MMS} interval used in CKH19 was chosen because it has a relative steady, constant magnitude magnetic field, as can be seen in Fig.~1 of Chen and Bolyrev (2017) \citep{Chen:2017}.  However, there are fluctuations in the direction of the magnetic field, and the projection of measurements onto a magnetic field-aligned coordinate system in CKH19 uses only the mean magnetic field direction averaged over the full 70~s interval.  Thus, changes in the direction of the instantaneous magnetic field relative to the fixed  $(v_\parallel,v_\perp)$ coordinate system could lead to some smearing out of the structures in  $C'_{E_\parallel, e}(v_\parallel,v_\perp)$. The \T{AstroGK} simulation, in contrast, employs a gyrotropic velocity space $(v_\parallel,v_\perp)$ coordinate system that is always aligned with the local magnetic field direction, enabling much finer structure to be resolved in the field-particle correlations. Such an effect of fluctuations in the instantaneous magnetic field direction leading to broadening of the resulting velocity-space signatures has indeed been found numerically in an analysis of ion Landau damping signatures in a hybrid Vlasov-Maxwell simulation of ion-cyclotron turbulence  using the \T{HVM} code\citep{Valentini:2007}, in which the velocity-space coordinate system does not follow the magnetic field \citep{Klein:2020}.

\subsection{Timestack Plots and Long-time Average of $ C_{E_\parallel, e}(v_{\parallel})$ }

The observational signature of electron Landau damping was found to be persistent throughout the entire 70~s \emph{MMS} data interval when the correlation was applied to a set of 10 equal-time subintervals \citep{Chen:2019}. In the simulation, we find examples of both persistent signatures and transient signatures, as shown in Fig.~\ref{fig:heatstacks}. Each of the twenty-four probe points showed evidence of at least transient signatures, and sixteen of the total contained at least one long-duration coherent signature for a minimum of approximately $3/4$ of the simulation time. Of those sixteen, five contained at least one signature that persisted throughout the entire simulation interval. This type of spatial variation is to be expected in kinetic turbulence \citep{TenBarge:2013a,Howes:2018}. Furthermore, the same sliding time-average window of $\tau/T_0=4$ driving-scale Alfv\'en waves was used in the timestack correlations of the simulation, which is again a much smaller correlation interval than what was used in the analogous plot in CKH19 (Fig 2(b) of that work), which corresponds to 28 wave periods at our domain scale. It is possible that many transient features that may otherwise have been visible would be obscured by such a long interval, leaving only the dominant signature about the electron thermal velocity to be observed. 

Another notable difference between our simulations and the observations is the lack of symmetric velocity-space signatures about $v_\parallel = 0$ in our simulation results. We do observe at least one instance of a symmetric pair of signatures (probe 15); however, it is not a ubiquitous scenario, at least under these averaging and sampling conditions, and it is seen to be quite transient. To enable a more direct comparison between the two results, we considered the pseudo long-time average of the correlation by averaging all twenty-four of the time-integrated correlations $ C_{E_{\parallel, e}}(v_\parallel)$, and display the result in Fig.~\ref{fig:avg_intC}. For $v_{\parallel}>0$, this plot shows a dominant energy transfer from fields to electrons, with a zero-crossing of the bipolar signature near to the electron thermal velocity, which is qualitatively similar to the observational \emph{MMS} interval (found in  Fig. 2(c) of CKH19). Note that in the observational result, the resonance was found to be slightly above the electron thermal velocity and symmetric about $v_{\parallel}=0$. For waves propagating down the magnetic field ($v_{\parallel} < 0 $), a single clear signature is not produced in the pseudo long-time integration of $C_{E_\parallel, e}(v_\parallel)$. Instead, there appear to be two weaker signatures, one on either side of the electron thermal velocity, and the net energy transfer to electrons by KAWs propagating down the magnetic field has a much smaller magnitude than the transfer due to the upward propagating waves. In fact, most of our simulation probe points returned examples of overall imbalanced signatures between $v_\parallel <0$ and $v_{\parallel}>0$, and showed signatures with resonant crossings at multiple values of $v_\parallel$. It is possible that the result of the long-time average would have differed had our $z=0$ plane probe points been positioned as we originally intended. If this had been the case, we could put greater confidence in our assumption that the wave modes at each probe were spatially independent, but it is unclear how exactly this would have affected our results. 

Overall, our numerical results indicate that electron Landau damping in a turbulent plasma does not typically have a broad and symmetric velocity-space signature, as obtained in CKH19. Rather, the velocity-space signature of $C_{E_\parallel, e}(v_\parallel)$ may have both persistent and transient features, where the dispersive nature of KAWs leads to bipolar signatures of Landau damping that can appear anywhere over a range of parallel velocities that satisfy the resonance condition $v_\parallel = \omega/k_\parallel$. Additionally, a new study that has performed field-particle correlation analysis of electron energization in 20 different \emph{MMS} burst-mode intervals has shown that indeed the CKH19 result of a broad and symmetric signature is, in fact, atypical \citep{Afshari:2020}. In that study, asymmetric signatures about $v_\parallel=0$ were extremely common, more structured parallel correlations with multiple bipolar signatures at different parallel velocities were found, and only one quarter of the sample returned a symmetric signature. Thus, it appears that the numerical simulation results presented here, although not showing a clear qualitative agreement to the single CKH19 interval, are in fact a more general representation of the variety of signatures of electron Landau damping that can be found in a turbulent plasma.

\section{\label{sec:C} Conclusion }

In this investigation, we performed a gyrokinetic simulation of turbulence in the Earth's magnetosheath corresponding to the \emph{MMS} observation by Chen et al.~(2019), (CKH19)\citep{Chen:2019}, which, when analyzed using the field-particle correlation technique, yielded the first direct evidence of electron Landau damping in a turbulent space plasma. We aimed first to determine numerically the typical velocity-space signature of electron Landau damping in a turbulent plasma, with a particular goal of understanding how the dispersive nature of kinetic \Alfven waves in the dissipation range of plasma turbulence affects the bipolar appearance of the characteristic velocity-space signature of Landau damping, as discussed in Sec.~\ref{sec:diagram}. Second, we aimed to compare these numerical results to the velocity-space signature found in CKH19 in order to understand and interpret that result more thoroughly.

Our driven, strong KAW turbulence simulation produced a magnetic energy spectrum that is consistent with previous simulations of the dissipation range of solar wind turbulence \citep{Howes:2011,TenBarge:2013a,TenBarge:2013b} and with the \emph{MMS} observations in our target interval \citep{Chen:2017,Chen:2019}, and the dissipation of the turbulent energy in steady-state led to heating of the electrons. We find that the time-averaged value of the work done by the parallel electric field $\langle j_\parallel  E_\parallel \rangle_\tau$ is largely positive over the simulation domain, occurs on scales significantly smaller than the driving scale of the simulation, and dominates the electron energization.  All of these features are consistent with dissipation dominated by electron Landau damping. 

The application of the field-particle correlation technique yielded gyrotropic velocity-space signatures $C_{E_\parallel}(v_\parallel,v_\perp)$ that showed the characteristic bipolar signatures of Landau damping at multiple parallel velocities over the range expected for the Landau resonance of dispersive kinetic \Alfven waves for the plasma parameters of this interval. When integrated over $v_\perp$ to yield a reduced parallel velocity-space signature as a function of time, $C_{E_\parallel}(v_\parallel,t)$, some of these bipolar features were found to be persistent over the full duration of the simulation, while others transiently appeared and disappeared. The time-integrated parallel velocity-space signatures, $C_{E_\parallel}(v_\parallel)$, showed multiple bipolar resonant zero-crossings, indicative of electron Landau damping of KAWs with particular values of $k_\perp \rho_i$, as well as broadened signatures that may be the result of a superposition of bipolar signatures of damped KAWs with nearby values of resonant parallel velocities, as illustrated in Fig.~\ref{fig:LDdiagramFull}. Together, these results suggest that electron Landau damping does not typically generate the clean signatures of ion Landau damping that occur at the relatively non-dispersive ion scales. Instead, the velocity-space signature of electron Landau damping in $C_{E_\parallel}(v_\parallel)$ may have both persistent and transient features, where the dispersive nature of KAWs leads to bipolar signatures of resonant Landau damping that can appear anywhere over a range of parallel velocities satisfying the resonance condition $v_\parallel = \omega/k_\parallel$. Indeed, some of those bipolar signatures can overlap in $v_\parallel$, leading to a broadened appearance of the signature. 

Comparing to the observational results, we do not typically find the symmetric and smooth appearance of the bipolar velocity-space signatures seen in CKH19. We do find instances of symmetric signatures, but they are not as common as the more varying signatures shown in Fig.~\ref{fig:heatstacks}. The lack of smoothness in our numerical results could be due to the higher velocity-space resolution in the simulations, the smearing effect of fluctuations in the magnetic field direction of the observation, or the much longer period over which the observations are time-averaged.  Averaging the reduced parallel correlation from all twenty-four of our probes to yield a longer effective time average, we present in Fig.~\ref{fig:avg_intC} the best comparison of our numerical $C_{E_\parallel}(v_\parallel)$ to the observational velocity-space signature in Fig.~2(c) of CKH19. Although we do not recover a signature that is symmetric about $v_\parallel=0$ as in the observations, the broadened bipolar signature on the $v_\parallel>0$ half is qualitatively consistent with the bipolar signature at $v_\parallel>0$ in the observations. Furthermore, a new field-particle correlation analysis of 20 different \emph{MMS} burst-mode intervals has shown that indeed the CKH19 result of a broad and symmetric signature is, in fact, atypical \citep{Afshari:2020}, and that the varied appearance of the velocity-space signatures of electron Landau damping in our numerical results may indeed be much more representative of the results from a larger sample of observations. Taken together, we believe that our simulation results provide a numerical confirmation that the results in CKH19 indeed provide evidence of electron Landau damping in a turbulent space plasma. Future numerical simulations will determine the characteristic velocity-space signature of electron Landau damping in the dispersive kinetic \Alfven wave regime over a wider range of plasma parameters that are relevant to other heliospheric environments of interest, such as the inner heliosphere.

\begin{acknowledgments}
This work was supported by NASA grants 80NSSC18K1371, 80NSSC18K0643, and 80NSSC18K1217 and NSF grant AGS-1842561.  This work used the Extreme Science and Engineering Discovery Environment (XSEDE), which is supported by National Science Foundation grant number ACI-1548562, on Stampede2 at the Texas Advanced Computing Center through NSF XSEDE Award TG-PHY090084.
\end{acknowledgments}
The data that support the findings of this study are available from the corresponding author upon reasonable request.

\appendix
\section{Linear Gyrokinetic Dispersion Relation}
\label{app:lindisp}
\begin{figure}[h]
\vspace{-0.9in}
\includegraphics[width=0.5\textwidth, height=0.5\textwidth]{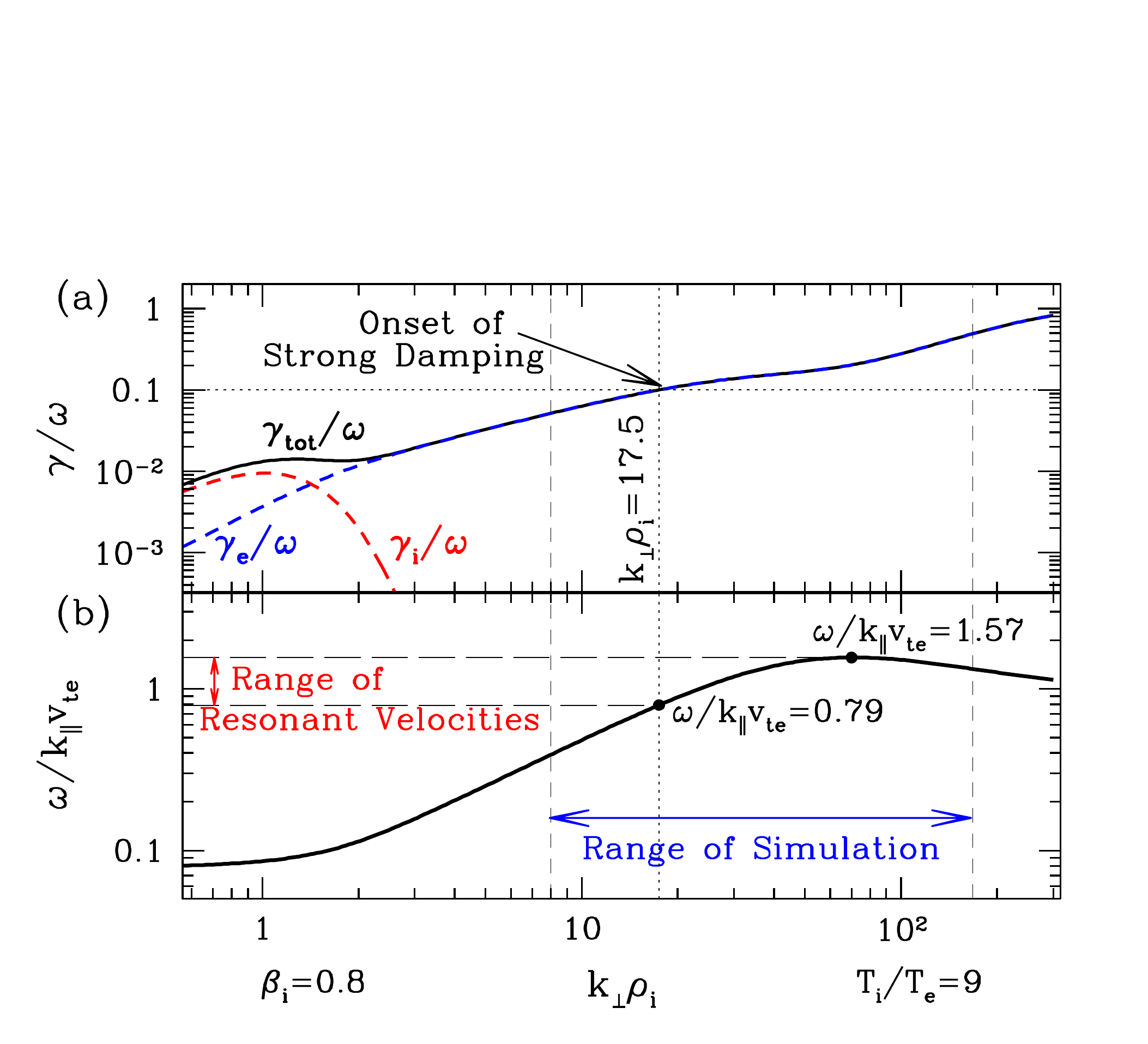}%
\caption{\label{fig:LGKDR} Linear gyrokinetic dispersion relation: (a) the ratio of total damping rate to frequency $\gamma_{tot}/\omega$ (black) and the separate ion  contribution $\gamma_i/\omega$ (red dashed) and electron  contribution $\gamma_e/\omega$ (blue dashed) vs.~perpendicular wavenumber, $k_\perp \rho_i$; (b) the normalized real frequency $\omega /k_\parallel v_{te}$ vs.~$k_\perp \rho_i$.}
\end{figure}

Here we present the linear gyrokinetic dispersion relation over the range of perpendicular wavenumbers resolved within the simulation. In Fig.~\ref{fig:LGKDR}(a), we plot the ratio of the damping rate to the frequency ($\gamma/\omega$) and in (b) the dimensionless frequency ($\omega /k_\parallel v_{te}$), which is the phase velocity normalized by the electron thermal speed. Both are plotted against the dimensionless quantity $k_\perp \rho_i$. Note that for perpendicular length scales larger than the ion Larmor radius, $k_\perp \rho_i < 1 $, which is the range of scales in the inertial range of the turbulence, the MHD Alfv\'en waves are nondispersive (here $\omega /k_\parallel v_{te}$ independent of $k_\perp \rho_i$). For perpendicular scales smaller than the ion Larmor radius, $k_\perp \rho_i > 1$, the MHD \Alfven wave transitions to the kinetic \Alfven wave, a dispersive wave mode where  $\omega /k_\parallel v_{te}$ increases approximately linearly with $k_\perp \rho_i$ until the waves approach the electron scales at $k_\perp \rho_e  \sim 1$ (equivalent to $k_\perp \rho_i \sim 128$). Note that these linear gyrokinetic dispersion relation results have been validated against the linear Vlasov-Maxwell dispersion relation in the limit $k_\perp \gg k_\parallel$. The dispersion relation in  Fig.~\ref{fig:LGKDR} clearly reveals the dispersive nature of kinetic Alfv\'en waves at $k_\perp\rho_i \gtrsim 1$, and is used below to show how the range of resonant velocities is determined. 

In panel (a) the total damping rate $\gamma_{tot}/\omega$ (black) is plotted along with the separate of ion (red dashed) and electron (blue dashed) contributions: $\gamma_i/\omega$ peaks at the ion Larmor radius, and $\gamma_{tot}/\omega$ is dominated by $\gamma_e/\omega$ well before the simulation driving scale, which is marked by the dashed vertical line at $k_{\perp 0} \rho_i = 8$. The maximum fully resolved wavenumber is $k_\perp \rho_i = 168$. A rule of thumb is that collisionless damping becomes significant when $\gamma_{tot}/\omega \gtrsim 0.1$, which begins at a perpendicular wavenumber $k_\perp \rho_i=17.5$. That perpendicular wavenumber is used in panel (b) to find the minimum resonant velocity $\omega /k_\parallel v_{te}=0.79$ at which we expect electron Landau damping to begin to become significant. The maximum frequency in the dispersion relation $\omega /k_\parallel v_{t,e}=1.57$ occurs at $k_\perp \rho_i \approx 71$, which gives the upper bound on the resonant velocities expected to contribute to electron Landau damping. Therefore, the range of resonant parallel velocities at which we expect to observe signatures of electron Landau damping is $0.79 \le |v_{\parallel}/v_{t,e}| \le 1.6$.
 

\begin{thebibliography}{10}

\bibitem{Barnes:1966}
A.~{Barnes}.
\newblock {Collisionless Damping of Hydromagnetic Waves}.
\newblock {\em Phys.~Fluids}, 9(8):1483--1495, 1966.

\bibitem{TenBarge:2013a}
J.~M. {TenBarge} and G.~G. {Howes}.
\newblock {Current Sheets and Collisionless Damping in Kinetic Plasma
  Turbulence}.
\newblock {\em Astrophys.~J.~Lett.}, 771(2):L27, Jul 2013.

\bibitem{Schekochihin:2009}
A.~A. {Schekochihin}, S.~C. {Cowley}, W.~{Dorland}, G.~W. {Hammett}, G.~G.
  {Howes}, E.~{Quataert}, and T.~{Tatsuno}.
\newblock {Astrophysical Gyrokinetics: Kinetic and Fluid Turbulent Cascades in
  Magnetized Weakly Collisional Plasmas}.
\newblock {\em Astrophys.~J.}, 182:310--377, May 2009.

\bibitem{Chen:2001}
Liu Chen, Zhihong Lin, and Roscoe White.
\newblock On resonant heating below the cyclotron frequency.
\newblock {\em Phys.~Plasmas}, 8(11):4713--4716, 2001.

\bibitem{Lichko:2017}
E.~{Lichko}, J.~{Egedal}, W.~{Daughton}, and J.~{Kasper}.
\newblock {Magnetic Pumping as a Source of Particle Heating and Power-law
  Distributions in the Solar Wind}.
\newblock {\em Astrophys.~J.~Lett.}, 850:L28, December 2017.

\bibitem{Dmitruk:2004}
P.~{Dmitruk}, W.~H. {Matthaeus}, and N.~{Seenu}.
\newblock {Test Particle Energization by Current Sheets and Nonuniform Fields
  in Magnetohydrodynamic Turbulence}.
\newblock {\em Astrophys.~J.}, 617:667--679, December 2004.

\bibitem{Karimabadi:2013}
H.~Karimabadi, V.~Roytershteyn, M.~Wan, W.~H. Matthaeus, W.~Daughton, P.~Wu,
  M.~Shay, B.~Loring, J.~Borovsky, E.~Leonardis, S.~C. Chapman, and T.~K.~M.
  Nakamura.
\newblock Coherent structures, intermittent turbulence, and dissipation in
  high-temperature plasmas.
\newblock {\em Phys.~Plasmas}, 20(1):012303, 2013.

\bibitem{Howes:2006}
G.~G. {Howes}, S.~C. {Cowley}, W.~{Dorland}, G.~W. {Hammett}, E.~{Quataert},
  and A.~A. {Schekochihin}.
\newblock {Astrophysical Gyrokinetics: Basic Equations and Linear Theory}.
\newblock {\em Astrophys.~J.}, 651:590--614, November 2006.

\bibitem{Tatsuno:2009}
T.~{Tatsuno}, W.~{Dorland}, A.~A. {Schekochihin}, G.~G. {Plunk}, M.~{Barnes},
  S.~C. {Cowley}, and G.~G. {Howes}.
\newblock {Nonlinear Phase Mixing and Phase-Space Cascade of Entropy in
  Gyrokinetic Plasma Turbulence}.
\newblock {\em Phys.~Rev.~Lett.}, 103(1):015003, July 2009.

\bibitem{Howes:2017}
G.~G. {Howes}, K.~G. {Klein}, and T.~C. {Li}.
\newblock {Diagnosing collisionless energy transfer using field-particle
  correlations: Vlasov-Poisson plasmas}.
\newblock {\em J.~Plasma Phys.}, 83(1):705830102, February 2017.

\bibitem{Klein:2017}
K.~G. {Klein}, G.~G. {Howes}, and J.~M. {Tenbarge}.
\newblock {Diagnosing collisionless energy transfer using field-particle
  correlations: gyrokinetic turbulence}.
\newblock {\em J.~Plasma Phys.}, 83(4):535830401, August 2017.

\bibitem{Howes:2017b}
Gregory~G. {Howes}.
\newblock {A prospectus on kinetic heliophysics}.
\newblock {\em Phys.~Plasmas}, 24(5):055907, May 2017.

\bibitem{Howes:2018}
Gregory~G. Howes, Andrew~J. McCubbin, and Kristopher~G. Klein.
\newblock Spatially localized particle energization by landau damping in
  current sheets produced by strong alfvén wave collisions.
\newblock {\em Journal of Plasma Physics}, 84(1):905840105, 2018.

\bibitem{Klein:2020}
K.~G. {Klein}, G.~G. {Howes}, J.~M. {TenBarge}, and F.~{Valentini}.
\newblock Diagnosing collisionless energy transfer using field-particle
  correlations: Alfv\'en-ion cyclotron turbulence.
\newblock {\em J.~Plasma Phys.}, 2020.
\newblock in press.

\bibitem{Howes:2008b}
G.~G. {Howes}, S.~C. {Cowley}, W.~{Dorland}, G.~W. {Hammett}, E.~{Quataert},
  and A.~A. {Schekochihin}.
\newblock {A model of turbulence in magnetized plasmas: Implications for the
  dissipation range in the solar wind}.
\newblock {\em J.~Geophys.~Res.}, 113(A5):A05103, May 2008.

\bibitem{Howes:2015}
G.~G. {Howes}.
\newblock {A dynamical model of plasma turbulence in the solar wind}.
\newblock {\em Philosophical Transactions of the Royal Society of London Series
  A}, 373(2041):20140145--20140145, Apr 2015.

\bibitem{Kiyani:2015}
K.~H. {Kiyani}, K.~T. {Osman}, and S.~C. {Chapman}.
\newblock {Dissipation and heating in solar wind turbulence: from the macro to
  the micro and back again}.
\newblock {\em Phil. Trans. R. Soc. A}, 373(2041):20140155--20140155, April
  2015.

\bibitem{Howes:2011b}
G.~G. {Howes}, J.~M. {Tenbarge}, and W.~{Dorland}.
\newblock {A weakened cascade model for turbulence in astrophysical plasmas}.
\newblock {\em Phys.~Plasmas}, 18(10):102305--102305, October 2011.

\bibitem{Chen:2019}
C.~H.~K. {Chen}, K.~G. {Klein}, and G.~G. {Howes}.
\newblock {Evidence for electron Landau damping in space plasma turbulence}.
\newblock {\em Nature Communications}, 10:740, February 2019.

\bibitem{Klein:2016}
K.~G. {Klein} and G.~G. {Howes}.
\newblock {Measuring Collisionless Damping in Heliospheric Plasmas using
  Field-Particle Correlations}.
\newblock {\em Astrophys.~J.~Lett.}, 826:L30, August 2016.

\bibitem{Howes:2008c}
Gregory~G. {Howes}.
\newblock {Inertial range turbulence in kinetic plasmas}.
\newblock {\em Phys.~Plasmas}, 15(5):055904--055904, May 2008.

\bibitem{Howes:2015b}
Gregory~G. {Howes}.
\newblock {\em {Kinetic Turbulence}}, volume 407 of {\em Astrophysics and Space
  Science Library}, page 123.
\newblock 2015.

\bibitem{Tu:1995}
C.-Y. {Tu} and E.~{Marsch}.
\newblock {MHD structures, waves and turbulence in the solar wind: Observations
  and theories}.
\newblock {\em Space Sci.~Rev.}, 73:1--2, July 1995.

\bibitem{Bruno:2005}
R.~{Bruno} and V.~{Carbone}.
\newblock {The Solar Wind as a Turbulence Laboratory}.
\newblock {\em Living Reviews in Solar Physics}, 2:4, September 2005.

\bibitem{Kolmogorov:1991}
A.~N. {Kolmogorov}.
\newblock {The Local Structure of Turbulence in Incompressible Viscous Fluid
  for Very Large Reynolds Numbers}.
\newblock {\em Proc. R. Soc. London, Ser. A}, 434(1890):9--13, July 1991.

\bibitem{Goldreich:1995}
P.~{Goldreich} and S.~{Sridhar}.
\newblock {Toward a Theory of Interstellar Turbulence. II. Strong Alfvenic
  Turbulence}.
\newblock {\em Astrophys.~J.}, 438:763, January 1995.

\bibitem{Boldyrev:2006}
Stanislav {Boldyrev}.
\newblock {Spectrum of Magnetohydrodynamic Turbulence}.
\newblock {\em Phys.~Rev.~Lett.}, 96(11):115002, March 2006.

\bibitem{Numata:2010}
R.~{Numata}, G.~G. {Howes}, T.~{Tatsuno}, M.~{Barnes}, and W.~{Dorland}.
\newblock {AstroGK: Astrophysical gyrokinetics code}.
\newblock {\em J.~Plasma Phys.}, 229:9347--9372, December 2010.

\bibitem{Howes:2008}
G.~G. {Howes}, W.~{Dorland}, S.~C. {Cowley}, G.~W. {Hammett}, E.~{Quataert},
  A.~A. {Schekochihin}, and T.~{Tatsuno}.
\newblock {Kinetic Simulations of Magnetized Turbulence in Astrophysical
  Plasmas}.
\newblock {\em Phys.~Rev.~Lett.}, 100(6):065004, February 2008.

\bibitem{Howes:2011}
G.~G. {Howes}, J.~M. {Tenbarge}, W.~{Dorland}, E.~{Quataert}, A.~A.
  {Schekochihin}, R.~{Numata}, and T.~{Tatsuno}.
\newblock {Gyrokinetic Simulations of Solar Wind Turbulence from Ion to
  Electron Scales}.
\newblock {\em Phys.~Rev.~Lett.}, 107(3):035004, July 2011.

\bibitem{TenBarge:2013b}
J.~M. {TenBarge}, G.~G. {Howes}, and W.~{Dorland}.
\newblock {Collisionless Damping at Electron Scales in Solar Wind Turbulence}.
\newblock {\em Astrophys.~J.}, 774:139, September 2013.

\bibitem{Cho:2002}
Jungyeon {Cho}, Alex {Lazarian}, and Ethan~T. {Vishniac}.
\newblock {New Regime of Magnetohydrodynamic Turbulence: Cascade below the
  Viscous Cutoff}.
\newblock {\em Astrophys.~J.~Lett.}, 566(1):L49--L52, February 2002.

\bibitem{Sahraoui:2010}
F.~{Sahraoui}, M.~L. {Goldstein}, G.~{Belmont}, P.~{Canu}, and L.~{Rezeau}.
\newblock {Three Dimensional Anisotropic k Spectra of Turbulence at Subproton
  Scales in the Solar Wind}.
\newblock {\em Phys.~Rev.~Lett.}, 105(13):131101, September 2010.

\bibitem{TenBarge:2014}
J.~M. {TenBarge}, G.~G. {Howes}, W.~{Dorland}, and G.~W. {Hammett}.
\newblock {An oscillating Langevin antenna for driving plasma turbulence
  simulations}.
\newblock {\em Comp.~Phys.~Comm.}, 185(2):578--589, February 2014.

\bibitem{Howes:2017c}
Gregory~G. {Howes} and Sofiane {Bourouaine}.
\newblock {The development of magnetic field line wander by plasma turbulence}.
\newblock {\em J.~Plasma Phys.}, 83(4):905830408, August 2017.

\bibitem{Alexandrova:2009}
O.~{Alexandrova}, J.~{Saur}, C.~{Lacombe}, A.~{Mangeney}, J.~{Mitchell}, S.~J.
  {Schwartz}, and P.~{Robert}.
\newblock {Universality of Solar-Wind Turbulent Spectrum from MHD to Electron
  Scales}.
\newblock {\em Phys.~Rev.~Lett.}, 103(16):165003, October 2009.

\bibitem{Alexandrova:2012}
O.~{Alexandrova}, C.~{Lacombe}, A.~{Mangeney}, R.~{Grappin}, and
  M.~{Maksimovic}.
\newblock {Solar Wind Turbulent Spectrum at Plasma Kinetic Scales}.
\newblock {\em Astrophys.~J.}, 760:121, December 2012.

\bibitem{Sahraoui:2013b}
F.~{Sahraoui}, S.~Y. {Huang}, G.~{Belmont}, M.~L. {Goldstein}, A.~{R{\'e}tino},
  P.~{Robert}, and J.~{De Patoul}.
\newblock {Scaling of the Electron Dissipation Range of Solar Wind Turbulence}.
\newblock {\em Astrophys.~J.}, 777:15, November 2013.

\bibitem{Zhdankin:2015a}
V.~{Zhdankin}, D.~A. {Uzdensky}, and S.~{Boldyrev}.
\newblock {Temporal Intermittency of Energy Dissipation in Magnetohydrodynamic
  Turbulence}.
\newblock {\em Phys.~Rev.~Lett.}, 114(6):065002, February 2015.

\bibitem{Zhdankin:2015b}
V.~{Zhdankin}, D.~A. {Uzdensky}, and S.~{Boldyrev}.
\newblock {Temporal Analysis of Dissipative Structures in Magnetohydrodynamic
  Turbulence}.
\newblock {\em Astrophys.~J.}, 811:6, September 2015.

\bibitem{Zhdankin:2016}
V.~{Zhdankin}, S.~{Boldyrev}, and D.~A. {Uzdensky}.
\newblock {Scalings of intermittent structures in magnetohydrodynamic
  turbulence}.
\newblock {\em Phys.~Plasmas}, 23(5):055705, May 2016.

\bibitem{Mallet:2019}
Alfred {Mallet}, Kristopher~G. {Klein}, Benjamin D.~G. {Chand ran}, Daniel
  {Gro{\v{s}}elj}, Ian~W. {Hoppock}, Trevor~A. {Bowen}, Chadi~S. {Salem}, and
  Stuart~D. {Bale}.
\newblock {Interplay between intermittency and dissipation in collisionless
  plasma turbulence}.
\newblock {\em J.~Plasma Phys.}, 85(3):175850302, June 2019.

\bibitem{Taylor:1938}
G.~I. {Taylor}.
\newblock {The Spectrum of Turbulence}.
\newblock {\em {Proc. Roy. Soc. A}}, 164:476--490, 1938.

\bibitem{Howes:2014a}
G.~G. {Howes}, K.~G. {Klein}, and J.~M. {TenBarge}.
\newblock {Validity of the Taylor Hypothesis for Linear Kinetic Waves in the
  Weakly Collisional Solar Wind}.
\newblock {\em Astrophys.~J.}, 789:106, July 2014.

\bibitem{Zhdankin:2014}
V.~{Zhdankin}, S.~{Boldyrev}, J.~C. {Perez}, and S.~M. {Tobias}.
\newblock {Energy Dissipation in Magnetohydrodynamic Turbulence: Coherent
  Structures or ''Nanoflares''?}
\newblock {\em Astrophys.~J.}, 795:127, November 2014.

\bibitem{Pollock:2016}
C.~{Pollock}, T.~{Moore}, A.~{Jacques}, J.~{Burch}, U.~{Gliese}, Y.~{Saito},
  T.~{Omoto}, L.~{Avanov}, A.~{Barrie}, V.~{Coffey}, J.~{Dorelli},
  D.~{Gershman}, B.~{Giles}, T.~{Rosnack}, C.~{Salo}, S.~{Yokota}, M.~{Adrian},
  C.~{Aoustin}, C.~{Auletti}, S.~{Aung}, V.~{Bigio}, N.~{Cao}, M.~{Chandler},
  D.~{Chornay}, K.~{Christian}, G.~{Clark}, G.~{Collinson}, T.~{Corris}, A.~{De
  Los Santos}, R.~{Devlin}, T.~{Diaz}, T.~{Dickerson}, C.~{Dickson},
  A.~{Diekmann}, F.~{Diggs}, C.~{Duncan}, A.~{Figueroa-Vinas}, C.~{Firman},
  M.~{Freeman}, N.~{Galassi}, K.~{Garcia}, G.~{Goodhart}, D.~{Guererro},
  J.~{Hageman}, J.~{Hanley}, E.~{Hemminger}, M.~{Holland}, M.~{Hutchins},
  T.~{James}, W.~{Jones}, S.~{Kreisler}, J.~{Kujawski}, V.~{Lavu}, J.~{Lobell},
  E.~{LeCompte}, A.~{Lukemire}, E.~{MacDonald}, A.~{Mariano}, T.~{Mukai},
  K.~{Narayanan}, Q.~{Nguyan}, M.~{Onizuka}, W.~{Paterson}, S.~{Persyn},
  B.~{Piepgrass}, F.~{Cheney}, A.~{Rager}, T.~{Raghuram}, A.~{Ramil},
  L.~{Reichenthal}, H.~{Rodriguez}, J.~{Rouzaud}, A.~{Rucker}, Y.~{Saito},
  M.~{Samara}, J.-A. {Sauvaud}, D.~{Schuster}, M.~{Shappirio}, K.~{Shelton},
  D.~{Sher}, D.~{Smith}, K.~{Smith}, S.~{Smith}, D.~{Steinfeld},
  R.~{Szymkiewicz}, K.~{Tanimoto}, J.~{Taylor}, C.~{Tucker}, K.~{Tull},
  A.~{Uhl}, J.~{Vloet}, P.~{Walpole}, S.~{Weidner}, D.~{White}, G.~{Winkert},
  P.-S. {Yeh}, and M.~{Zeuch}.
\newblock {Fast Plasma Investigation for Magnetospheric Multiscale}.
\newblock {\em Space Sci.~Rev.}, 199:331--406, March 2016.

\bibitem{Chen:2017}
C.~H.~K. {Chen} and S.~{Boldyrev}.
\newblock {Nature of Kinetic Scale Turbulence in the Eath's Magnetosheath}.
\newblock {\em Astrophys.~J.}, 842:122, June 2017.

\bibitem{Valentini:2007}
F.~{Valentini}, P.~{Tr{\'a}vn{\'{\i}}{\v c}ek}, F.~{Califano}, P.~{Hellinger},
  and A.~{Mangeney}.
\newblock {A hybrid-Vlasov model based on the current advance method for the
  simulation of collisionless magnetized plasma}.
\newblock {\em J.~Comp.~Phys.}, 225:753--770, July 2007.

\bibitem{Afshari:2020}
A.~S. {Afshari}, G.~G. {Howes}, C.~A. {Kletzing}, D.~P. {Hartley}, and S.~A.
  {Boardsen}.
\newblock {Electron Landau Damping of Turbulence in the Terrestrial
  Magnetosheath Plasma}.
\newblock {\em Geophys.~Res.~Lett.}, 2020.
\newblock submitted.

\end{thebibliography}

\end{document}